\documentclass[12pt]{article}
\usepackage{amsmath}
\usepackage{amssymb}
\numberwithin{equation}{section}

\title{Pfaff $\tau$-functions}

\author{
M. Adler\thanks{ Department of Mathematics, Brandeis University,
Waltham, Mass 02454, USA. E-mail: adler@math.brandeis.edu.  The
support of a National Science Foundation grant
\# DMS-98-4-50790
 is
gratefully acknowledged.}~~~~~~ T. Shiota\thanks{ Department of
Mathematics, Kyoto University, Kyoto 606, Japan. The hospitality of
the University of Louvain and Brandeis University is gratefully
acknowledged.}~~~~~~ P. van Moerbeke\thanks{ Department of
Mathematics, Universit\'e de Louvain, 1348 Louvain-la-Neuve,
Belgium and Brandeis University, Waltham, Mass 02454, USA. E-mail:
vanmoerbeke@geom.ucl.ac.be and @math.brandeis.edu. The support of a
National Science Foundation grant \# DMS-98-4-50790, a Nato, a FNRS
and a Francqui Foundation grant is gratefully acknowledged.}}

\date{July 23, 1999}

\newcommand{\MAT}[1]{\left(\begin{array}{*#1c}}
\newcommand{\mat}{\end{array}\right)}
\newcommand{\qed}{\leavevmode\unskip\nobreak\penalty200\hskip2pt\null
\nobreak\hfill\rule{1.1ex}{1.1ex}\medbreak
}

\newcommand{\rg}{\rightarrow}

\newcommand{\TT}{\tilde\tau}

\newcommand{\LR}{{\cal L}}

\newcommand{\OR}{{\cal O}}

\newcommand{\BC}{{\mathbb C}}

\newcommand{\BX}{{\mathbb X}}
\newcommand{\BY}{{\mathbb Y}}
\newcommand{\BZ}{{\mathbb Z}}

\newcommand{\iy}{\infty}
\newcommand{\pl}{\partial}
\newcommand{\al}{\alpha}

\newenvironment{proof}{\medskip\noindent{\it Proof:\/} }{\qed}
\newenvironment{remark}
        {\medskip\noindent\underline{\it Remark:\/} }{\medbreak}

\newcommand{\dt}{\delta}
\newcommand{\Dt}{\Delta}

\newcommand{\BR}{{\mathbb R}}
\newcommand{\lb}{\lambda}
\newcommand{\Lb}{\Lambda}

\newcommand{\diag}{\operatorname{diag}}
\newcommand{\Res}{\operatorname{Res}}

\def\be#1\ee{\begin{equation}#1\end{equation}}
\def\bea#1\eea{\begin{eqnarray}#1\end{eqnarray}}
\def\bean#1\eean{\begin{eqnarray*}#1\end{eqnarray*}}

\newcommand{\Pf}{\operatorname{\rm Pfaff}}
\newcommand{\Tr}{\operatorname{\rm Tr}}
\newcommand{\Mat}{\operatorname{\rm Mat}}
\newcommand{\OneOrTwo}{{\left\{\substack{1\\2}\right\}}}

\newtheorem{definition}{Definition}[section]
\newtheorem{theorem}[definition]{Theorem}
\newtheorem{lemma}[definition]{Lemma}
\newtheorem{corollary}[definition]{Corollary}

\makeatletter
\def\ps@X{\let\@mkboth\@gobbletwo
        \def\@oddhead{\tt Adler-Shiota-van\,Moerbeke:%
        Pfaff $\tau$\hfil
        July 23, 1999
        \hfil\S\thesection, p.\thepage}
        \def\@oddfoot{\rm\hfil\thepage\hfil}
        \let\@evenhead\@oddhead
        \let\@evenfoot\@oddfoot}
\makeatother

\begin{document}
\maketitle


\tableofcontents

\setcounter{section}{-1} 
\section{Introduction}

Throughout, let $A:=\BZ$ (``bi-infinite'' case) or
$A:=\BZ_{\ge0}=\{0,1,\dots\}$ (``semi-infinite'' case).
Consider the set of equations
\begin{equation}
\frac{\pl m_\iy}{\pl t_n}=\Lb^n m_\iy,\quad\quad\quad
\frac{\pl m_\iy}{\pl s_n}=-m_\iy(\Lb^\top)^n,\quad n=1,2,\dots,
\label{0.1}
\end{equation}
on bi- or semi-infinite (i.e., $A\times A$) matrices $m_\iy=m_\iy(t,s)$,
where the matrix
$\Lb=(\dt_{i,j-1})_{i,j\in A}$ is the shift matrix, and $\Lb^\top$ its
transpose. In
\cite{AvM2,AvM4}, it was shown that Borel decomposing\footnote{
        Here ``$t$, $s\in\BC^\iy$'' is an informal way of saying that
        $t$ and $s$ are two sequences of independent scalar variables;
        a function of those variables may be defined only in an open subset
        of $\BC^\iy\times\BC^\iy$, or may even be a formal power series in
        $t$ and $s$.
}
\begin{equation}
m_\iy(t,s)=\left(\mu_{ij}\right)_{i,j\in A}=S_1^{-1}S_2,\quad
\mbox{ for ``generic'' $t$, $s\in\BC^\iy$},
\label{0.2}
\end{equation}
into lower- and upper-triangular matrices $S_1(t,s)$ and
$S_2(t,s)$, leads to a two-Toda (two-dimensional Toda) system for
$L_1:=S_1\Lb S_1^{-1}$ and $L_2=S_2\Lb^\top S_2^{-1}$,
with $\tau$-functions given\footnote{
        \label{fn2}
        This formula will be used mainly in the semi-finite case, with
        $m_n=(\mu_{ij})_{0\leq i,j<n}$.  In the bi-infinite case,
        $m_n=(\mu_{ij})_{-\iy<i,j<n}$ and the determinant is interpreted as
        $$
        \lim_{k\to\iy}\det(\mu_{ij})_{-2k\leq i,j<q},\eqno(*)
        $$
        assuming the limit makes sense.}
by
\begin{equation}
\tau_n(t,s)=\det m_n(t,s),\quad n\in A.
\label{0.3}
\end{equation}
This paper deals with {\em skew-symmetric\/} initial data $m_\iy(0,0)$.
As readily seen from formula (0.1), the 2-Toda flow then maintains the
relation $m_\iy(t,s)=-m_\iy(-s,-t)^\top$, and hence, by formula (\ref{0.3})
and the interpretation of its right hand side in footnote~\ref{fn2},
\begin{equation}
\tau_n(t,s)=(-1)^n\tau_n(-s,-t).
\label{0.4}
\end{equation}

The main point of this paper is to study the reduction $s=-t$, as
used in the theory of random matrices in H. Peng's doctoral
dissertation \cite{P}. When $s\to-t$, formula (\ref{0.4}) shows that
in the limit the odd $\tau$-functions
vanish, whereas the even $\tau$-functions are determinants of
skew-symmetric matrices. In particular, the factorization (\ref{0.2})
fails; in fact in the limit $s\to
-t$, the system leaves the main stratum to penetrate a {\em deeper
stratum\/} in the Borel decomposition. In this paper we show this
stratum leads to its own {\em system}, whereas in a forthcoming
paper with Horozov \cite{AHvM}, we show this system is {\em
integrable\/} by producing its Lax pair.

Thus, we are led to considering {\em Pfaffians\/}:
\begin{equation}
\TT_n(t):=\Pf m_n(t,-t)=
(\det m_n(t,-t))^{1/2}=
\tau_n(t,-t)^{1/2},
\label{0.5}
\end{equation}
for every {\em even\/} $n\in A$ (the same remark as in
footnote~\ref{fn2} applies here).
 The {\em ``Pfaffian $\TT$-function''\/} is itself {\em not\/} a
2-Toda $\tau$-function, but it ties up remarkably with the 2-Toda
$\tau$-function $\tau$ as follows\footnote{
        $[\al]:=(\al,\frac{\al^2}{2}, \frac{\al^3}{3},\dots)$}:
\begin{equation}
\begin{aligned}
\tau_{2n}(t,-t-[\al]+[\beta])&=
\TT_{2n}(t)\TT_{2n}(t+[\al]-[\beta])
\\
\tau_{2n+1}(t,-t-[\al]+[\beta])&=
(\beta-\al)\TT_{2n}(t-[\beta])\TT_{2n+2}(t+[\al]).
\end{aligned}
\label{0.6}
\end{equation}
When $\beta\to\al$, we approach the {\em deeper\/} stratum in the
Borel decomposition of $m_\iy$ in a very specific way. It also shows that
the odd $\tau$-functions $\tau_{2n+1}(t,-t-[\al]+[\beta])$ approach
zero linearly as $\beta\to\al$, at the rate depending on $\al$:
$$
\lim_{\beta\to\al}\tau_{2n+1}(t,-t-[\al]+[\beta])/(\beta-\al)
=\TT_{2n}(t-[\al])\TT_{2n+2}(t+[\al]).
$$
Equations (\ref{0.6}) are crucial in establishing
 bilinear relations\footnote{
        $\tilde\pl=(
        \frac{\pl}{\pl t_1},\frac{1}{2}\frac{\pl}{\pl t_2},
        \frac{1}{3}\frac{\pl}{\pl t_3},\dots)$, and
       $\tilde D=\left(
       D_1,\frac{1}{2}D_2,\frac{1}{3}D_3,\dots\right)$ is the corresponding
       Hirota symbol, i.e.,
        $P(\tilde D)f\cdot g
        := P(\pl/\pl y_1,\frac{1}{2}\pl/\pl y_2,\frac{1}{3}\pl/\pl y_3,\dots)
        f(t+y)g(t-y)|_{y=0}$ for any polynomial $P$; and $p_k$ are the
        elementary Schur functions:
        $\sum_0^\infty p_k(t)z^k:=\exp(\sum_1^\infty t_iz^i)$.
} for Pfaffian $\TT$-functions, where $n$,~$m\in A$:
\begin{multline}
\sum_{\substack{j,k\ge0\\j-k=-2n+2m+1}}
p_j(-2y) e^{\sum-y_iD_i}p_k(-\tilde D)\TT_{2n}\cdot\TT_{2m+2}\\
 {}+
\sum_{\substack{j,k\ge0\\k-j=-2n+2m-1}}
p_j(2y) e^{\sum-y_iD_i}p_k(\tilde D)\TT_{2n+2}\cdot\TT_{2m}
=0.
\label{0.8}
\end{multline}
This is a generating function for the Hirota equations satisfied by $\TT(t)$:
for each $n$ and $m$, after expanding (\ref{0.8}) into a power series in
$y=(y_1,y_2,\dots)$, the coefficient of each monomial in $y$ gives a Hirota
equation.  For example, for $m=n-1$, the coefficient of the linear terms
($y_{k+3}$, to be more specific) gives
\begin{equation}
\biggl(p_{k+4}(-\tilde D)-\frac{1}{2}D_1D_{k+3}\biggr)\TT_{2n}\cdot\TT_{2n}
=p_k(\tilde D)\TT_{2n+2}\cdot\TT_{2n-2},
\label{0.9}
\end{equation}
where $k\in\BZ_{\ge0}$ and $2n-2\in A$. For $k=0$, this equation
can be viewed as an inductive expression of $\TT_{2n+2}$ in terms
of $\TT_{2n-2}$ and derivatives of $\TT_{2n}$. These equations
already appear in the work of Kac and van de Leur \cite{KvdL}, in
the context of the DKP hierarchy. On the exact connection, see
forthcoming work by J. van de Leur \cite{vdL}.

In analogy with the 2-Toda or KP theory, one establishes Fay
identities for the Pfaff $\TT$-functions. In this instance, they
involve Pfaffians rather than determinants:

\begin{multline}
\Pf\biggl(
\frac{(z_j-z_i)\TT_{2n-2}(t-[z_i]-[z_j])}{\TT_{2n}(t)}
\biggr)_{1\le i,j\le 2k}
\\
=\Dt(z)\frac{\TT_{2n-2k}\Bigl(t-\sum_1^{2k}[z_i]\Bigr)}{\TT_{2n}(t)}.
 \label{PfaffFay}
\end{multline}

In the semi-infinite case, the latter has a useful interpretation
in terms of Pfaffians of ``Christoffel-Darboux" kernels of the form
\begin{equation}
K_n(\mu,\lb)= e^{\sum_1^\iy t_i(\mu^i+\lb^i)}
\sum_0^{n-1}\Bigl(
q_{2k}(t,\lb)q_{2k+1}(t,\mu) - q_{2k}(t,\mu)q_{2k+1}(t,\lb)
\Bigr),
\label{4.8}
\end{equation}
where the $q_m(t,\lambda)$ form a system of skew-orthogonal
polynomials \cite{AHvM}. This is the analogue of the
Christoffel-Darboux kernel for orthogonal polynomials. So, formula
(0.9) can be rewritten as

\begin{equation}
\Pf(K_n(z_i,z_j))_{1\le i,j\le 2k}=
\Biggl(\frac{1}{\TT}
 \prod_{\substack{i=1\\\text{ordered}}}^{2k}\BX(t;z_i)\TT
\Biggr)_{2n},
\label{4.6}
\end{equation}
where $\BX(t;z)$ is a vertex operator for the corresponding Pfaff
 lattice (see \cite{AHvM, AvM6}):
 $$
 \BX(t;z):=\Lb^{-1} e^{\sum_1^{\iy}t_i z^i}
e^{-\sum_1^{\iy}\frac{z^{-i}}{i}\frac{\pl}{\pl t_i}}\chi(z)
. $$
This vertex operator also has the remarkable property
 that for a Pfaffian $\TT$-function,
\bean
&&\hspace{-0.5cm}\TT+a\BX(\lb)\BX(\mu)\TT \\ &&
   \hspace{1cm}=\TT_{2n}(t)+ a
\left(1-\frac{\mu}{\lb}\right)\lb^{2n-2}\mu^{2n-1}e^
{\sum
t_{1}(\lb^i+\mu^i)}\TT_{2n-2}(t-[\lb^{-1}]-[\mu^{-1}]).
\eean
is again a Pfaffian $\TT$-function.

 As was shown in \cite{AvM2, AvM3}, the 2-Toda lattice has
four distinct vertex operators. Upon using the reduction $s=-t$,
the 2-Toda vertex operators reduce to vertex operators for the
Pfaff lattice. This enables us to give the action of Virasoro
generators on Pfaff $\TT$-functions, in terms of the restriction
(to $s=-t$) of actions on 2-Toda $\tau$-functions:
$$
\left(J_i^{(k)}(t)+(-1)^kJ_i^{(k)}(s)\right)\tau_{2n}(t,s)|_{s=-t}
=2\TT_{2n}(t)J_i^{(k)}(t)\TT_{2n}(t).
$$
Finally, we discuss two examples, a first sketchy one, involving a
{\em semi-infinite} Pfaff lattice and matrix integrals; this is
extensively discussed in \cite{AvM6}. A second example, genuinely
{\em bi-infinite}, will be given in the context of curves with
fixed point free involutions $\iota$, equipped with a line bundle
$\LR$ having a suitable antisymmetry condition with respect to
$\iota$.

\section{Borel decomposition and the 2-Toda lattice}

In \cite{AvM4,AvM2}, we considered the following differential
equations for the bi-infinite or semi-infinite moment matrix
$m_\iy$
\begin{equation}
\frac{\pl m_\iy}{\pl t_n}=\Lb^n m_\iy,\quad
\frac{\pl m_\iy}{\pl s_n}=-m_\iy(\Lb^\top)^n,\quad n=1,2,\dots,
\label{1.1}
\end{equation}
where the matrix $\Lb=(\dt_{i,j-1})_{i,j\in A}$ is the shift
matrix; then (\ref{1.1}) has the following solution
\begin{equation}
m_\iy(t,s)=e^{\sum t_n\Lb^n}m_\iy(0,0)e^{-\sum s_n(\Lb^\top)^n}
\label{1.2}
\end{equation}
in terms of the initial data $m_\iy(0,0)$.

Assume $m_\iy$ allows, for ``generic'' $(t,s)$,
the Borel decomposition $m_\iy=S_1^{-1}S_2$, for
\begin{align*}
S_1\in G_-&:=\biggl\{\begin{tabular}{@{}l@{}}
lower-triangular matrices\\with 1's on the diagonal
\end{tabular}\biggr\},\\
S_2\in G_+&:=\biggl\{\begin{tabular}{@{}l@{}}
upper-triangular matrices\\with non-zero diagonal entries
\end{tabular}\biggr\},
\end{align*}
with corresponding Lie algebras $g_-$, $g_+$. Assume moreover that
$m_\iy$, $S_1$ and $S_2$ are nice in the sense of Remark at the end
of this section.  Then setting $L_1:=S_1\Lb S_1^{-1}$,
$$
S_1\frac{\pl m_\iy}{\pl t_n}S_2^{-1}=\begin{cases}
S_1(\pl/\pl t_1)(S_1^{-1}S_2)S_2^{-1}=-\dot S_1S_1+\dot
S_2S_2^{-1}\in g_-+g_+,
\\[3pt]
S_1\Lb^n m_\iy
S_2^{-1}=S_1\Lb^nS_1^{-1}=L_1^n=(L_1^n)_-+(L_1^n)_+\in g_-+g_+;
\end{cases}
$$
the uniqueness of the decomposition $g_-+g_+$ leads to
$$
-\frac{\pl S_1}{\pl t_n}S_1^{-1}=(L_1^n)_-,\quad
\frac{\pl S_2}{\pl t_n}S_2^{-1}=(L_1^n)_+.
$$
Similarly, setting $L_2=S_2\Lb^\top S_2^{-1}$, we find
$$
-\frac{\pl S_1}{\pl s_n}S_1^{-1}=-(L_2^n)_-,\quad
\frac{\pl S_2}{\pl s_n}S_2^{-1}=-(L_2^n)_+.
$$
This leads to the 2-Toda equations \cite{UT} for $S_1$, $S_2$ and
$L_1$, $L_2$:
\begin{equation}
\frac{\pl}{\pl t_n}S_\OneOrTwo
=\mp(L_1^n)_\mp S_\OneOrTwo,\quad
\frac{\pl}{\pl s_n}S_\OneOrTwo=\pm(L_2^n)_\mp S_\OneOrTwo,
\label{1.3}
\end{equation}
\begin{equation}
\frac{\pl L_i}{\pl t_n}=[(L_1^n)_+,L_i],\quad
\frac{\pl L_i}{\pl s_n}=[(L_2^n)_-,L_i],\quad i=1,2,\dots,
\label{1.4}
\end{equation}
and conversely, reading this argument backwards, we observe that
the 2-Toda equations (\ref{1.3}) imply the time evolutions (\ref{1.1})
for $m_\iy$.

The pairs of wave and adjoint wave functions
$\Psi=(\Psi_1,\Psi_2)$ and $\Psi^*=(\Psi_1^*,\Psi_2^*)$,
defined by
\begin{equation}
\begin{aligned}
\Psi_\OneOrTwo(t,s,z)&=
e^{\sum_1^\iy\left\{\substack{t_i\\s_i}\right\}z^{\pm i}}
S_\OneOrTwo\chi(z),
\\
\Psi^*_\OneOrTwo(t,s,z)&=
e^{-\sum_1^\iy\left\{\substack{t_i\\s_i}\right\}z^{\pm i}}
\Bigl(S^\top_\OneOrTwo\Bigr)^{-1}\chi(z^{-1}),
\end{aligned}
\label{1.5}
\end{equation}
where $\chi(z)$ is the column vector $(z^n)_{n\in A}$,
satisfy
$$
L\Psi=(z,z^{-1})\Psi,\quad L^*\Psi^*=(z,z^{-1})\Psi^*,
$$
and
\begin{equation}
\begin{aligned}
\frac{\pl }{\pl t_n}\Psi&=((L_1^n)_+,(L_1^n)_+)\Psi,
\\
\frac{\pl }{\pl s_n}\Psi&=((L_2^n)_-,(L_2^n)_-)\Psi,
\\
\frac{\pl}{\pl t_n}\Psi^* &=-(((L_1^n)_+)^\top,((L_1^n)_+)^\top)\Psi^*,
\\
\frac{\pl}{\pl s_n}\Psi^* &=-(((L_2^n)_-)^\top,((L_2^n)_-)^\top)\Psi^*,
\end{aligned}
\label{1.6}
\end{equation}
which are equivalent to (\ref{1.3}), and are further equivalent to
the following bilinear identities,\footnote{
        The contour integral around $z=\iy$ is taken {\em clockwise\/} about
        a small circle around $z=\iy$, while the one around $z=0$ is taken
        counter-clockwise about $z=0$.}
for all $m,n\in A$ and $t,s,t',s'\in\BC^\iy$:
\begin{equation}
\oint_{z=\iy}\Psi_{1n}(t,s,z)\Psi_{1m}^*(t',s,',z')\frac{dz}{2\pi iz}
=\oint_{z=0} \Psi_{2n}(t,s,z) \Psi_{2m}^*(t',s,',z')\frac{dz}{2\pi iz}.
\label{1.7}
\end{equation}

By 2-Toda theory \cite{UT,AvM4}, the problem is solved in terms of a
sequence of tau-functions
\begin{equation}
\tau_n(t,s)=\det m_n(t,s),
\label{1.8}
\end{equation}
with $m_n(t,s)$ defined in (and $\det m_n$ interpreted as in)
footnote~\ref{fn2}:
\begin{equation}
m_n(t,s):=\begin{cases}
\bigl(\mu_{ij}(t,s)\bigr)_{-\iy<i,j<n}&\quad\mbox{(bi-infinite case)},
\\
\bigl(\mu_{ij}(t,s)\bigr)_{0\leq i,j<n},
&\quad\mbox{(semi-infinite case, with $\tau_0=1$)},
\end{cases}
\label{1.9}
\end{equation}
as
\begin{equation}
\begin{aligned}
\Psi_1(t,s;z)&=
\biggl(
        \frac{\tau_n(t-[z^{-1}],s)}{\tau_n(t,s)}
e^{\sum^\iy_1 t_iz^i}z^n
\biggr)_{n\in A}\,,
\\
\Psi_2(t,s;z)&=\biggl(
\frac{\tau_{n+1}(t,s-[z])}{\tau_n(t,s)}e^{\sum^\iy_1s_iz^{-i}}z^n
\biggr)_{n\in A}\,,
\\
\Psi^*_1(t,s,z)&=\biggl(
\frac{\tau_{n+1}(t+[z^{-1}],s)}{\tau_{n+1}(t,s)}e^{-\sum^\iy_1t_iz^i}z^{-n}
\biggr)_{n\in A}\,,
\\
\Psi^*_2(t,s,z)&=\biggl(
\frac{\tau_n(t,s+[z])}{\tau_{n+1}(t,s)}e^{-\sum^\iy_1 s_iz^{-i}}z^{-n}
\biggr)_{n\in A}\,.
\end{aligned}
\label{1.10}
\end{equation}
Note (\ref{1.5}) and (\ref{1.10}) yield
\begin{equation}
h(t,s):= \mbox{(diagonal part of $S_2$)} =
\diag\biggl(\frac{\tau_{n+1}(t,s)}{\tau_{n}(t,s)}\biggr)_{n\in A}.
\label{1.11}
\end{equation}

Formulas (\ref{1.7}) and (\ref{1.10}) imply the following bilinear identities
\begin{multline}
\oint_{z=\iy}\tau_n(t-[z^{-1}],s)\tau_{m+1}(t'+[z^{-1}],s')
e^{\sum_1^\iy(t_i-t'_i)z^i}
z^{n-m-1}dz
\\
=\oint_{z=0}\tau_{n+1}(t,s-[z])\tau_m(t',s'+[z])
e^{\sum_1^\iy(s_i-s'_i)z^{-i}}z^{n-m-1}dz,
\label{1.12}
\end{multline}
where $m$, $n\in A$,
satisfied by and characterizing the 2-Toda $\tau$-functions.

\begin{remark}

In the bi-infinite case the factorization $m_\iy=S_1^{-1}S_2$ in (\ref{0.2})
or the determinant formula (\ref{0.3}) may fail to make sense.
Nevertheless, we can take (\ref{0.4}) as a starting point, use (\ref{0.6})
to define $\TT$ up to the sign, and make sense of the $\tau$-side of the
whole story.

In the bi-infinite case, factorization as in (\ref{0.2}) is not unique
in general.
This is responsible for the Backlund transform of a finite band matrix
having a continuous family of solutions.  However, it means the
matrix multiplication may not be associative in the bi-infinite case: if
$$
S_1^{-1}S_2=S_1^{\prime-1}S_2',
$$
with
$S_1$,~$S_1'\in G_-:=\{$lower triangular matrices with 1's on the diagonal$\}$
and
$S_2$,~$S_2'\in G_+:=\{$upper triangular matrices with non-zero diagonal$\}$,
and if the matrix multiplication was always associative, we should have
$$
S_1'S_1^{-1}=S_2'S_2^{-1}\in G_+\cap G_-=\{1\},
$$
so that $S_1=S_1'$ and $S_2=S_2'$.

The associativity is important in establishing the relation between equation
(\ref{0.1}) and the 2-Toda flows on $(S_1,S_2)$.  Moreover, we are mainly
interested in the semi-infinite case, in which the associativity clearly
holds.  So we assume that, for generic $(t,s)$, $S_1$, $S_2$ and $m_\iy$
actually belong to suitable sub{\em groups\/} $G_\pm'$ of $G_\pm$ and
a suitable subspace $\Mat'$ of the space $\Mat$ of all infinite matrices,
respectively, in which the multiplications
$$
\begin{array}{ccc} G_-'\times\Mat'&\to&\Mat'\\(S,m)&\mapsto&Sm
\end{array}
\quad\mbox{and}\quad
\begin{array}{ccc} \Mat'\times G_+'&\to&\Mat'\\(m,S)&\mapsto&mS
\end{array}
$$
are associative:
$$
(SS')m=S(S'm),\quad m(S''S''')=(mS'')S'''\quad\mbox{and}\quad(Sm)S''=S(mS'')
$$
hold for any $S$,~$S'\in G_-'$, $S''$,~$S'''\in G_+'$ and $m\in\Mat'$.

For instance these conditions are satisfied if the $(i,j)$ entries of every
matrix in $G_\mp'$ and $\Mat'$ tend to 0 quickly enough as $i\to-\iy$ uniformly
in $j>i+a$, and as $j\to-\iy$ uniformly in $i>j+a$, for some constant $a$.

\end{remark}

\section{Two-Toda $\tau$-functions versus Pfaffian $\TT$-functions}

In this section, we assume either the matrix $m_\iy$ is semi-infinite,
or $\det m_n$ can be interpreted as in formula~$(*)$ in
footnote~\ref{fn2}, and we exhibit the properties of the 2-Toda lattice,
associated with a skew-symmetric initial matrix $m_\iy(0,0)$.
The $\tau$-functions $\tau_n(t,s)$ then have the property
$$
\tau_n(t,s)=(-1)^n\tau_n(-s,-t).
$$
\begin{theorem}
If the initial matrix $m_\iy(0,0)$ is skew-symmetric, then
under the 2-Toda flow, $m_\iy(t,s)$ maintains the relation
\begin{equation}
m_\iy(t,s)=-m_\iy(-s,-t)^\top.
\label{2.1}
\end{equation}
Moreover,
\begin{equation}
h^{-1}S_1(t,s)=-(S_2^\top)^{-1}(-s,-t),\quad
h^{-1}S_2(t,s)=(S_1^\top)^{-1}(-s,-t),
\label{2.2}
\end{equation}
\begin{equation}
h^{-1}\Psi_1(t,s,z)=-\Psi_2^*(-s,-t,z^{-1}),\quad
h^{-1}\Psi_2(t,s,z)=\Psi_1^*(-s,-t,z^{-1}),
\label{2.3}
\end{equation}
\begin{equation}
L_1(t,s)=h L_2^\top h^{-1}(-s,-t)\quad\mbox{and}\quad
L_2(t,s)=h L_1^\top h^{-1}(-s,-t),
\label{2.4}
\end{equation}
with $h$, defined by (\ref{1.11}), satisfying
\begin{equation}
h(-s,-t)=-h(t,s).
\label{2.5}
\end{equation}
Finally,
\begin{equation}
\tau_n(-s,-t)=(-1)^n\tau_n(t,s).
\label{2.6}
\end{equation}
\end{theorem}

\begin{proof}
Formula (\ref{2.1}) is an immediate consequence of (\ref{1.2}) and
the skew-symmetry of $m_\iy(0,0)$.
Formula (\ref{2.2}) follows from (\ref{2.1}) and the Borel decomposition
of $m_\iy(t,s)$ and $-m_\iy(-s,-t)^\top$:
\begin{align*}
m_\iy(t,s)&=S_1^{-1}(t,s)S_2(t,s),
\\[4pt]
-m_\iy(-s,-t)^\top&=-S_2^\top(-s,-t)S_1^{-1\top}(-s,-t)
\\
&=(S_2^\top(-s,-t)h^{-1}(-s,-t))(-h(-s,-t)S_1^{-1\top}(-s,-t)),
\end{align*}
by the uniqueness of the Borel decomposition of
$m_\iy(t,s)=-m_\iy(-s,-t)^\top$, we have
\begin{align*}
S_1^{-1}(t,s)&=S_2^\top(-s,-t)h^{-1}(-s,-t)\in G_-
\\
S_2(t,s)&=-h(-s,-t)S_1^{-1\top}(-s,-t)\in G_+ .
\end{align*}

Substituting $(t,s)\to (-s,-t)$ in the second equation and
comparing it to the first one, yields $h(t,s)=-h(-s,-t)$, which
is (\ref{2.5}). Substituting this relation into the first and second
equations yields (\ref{2.2}), which by (\ref{1.5}) and the definition of $L_1$
and $L_2$, amounts to (\ref{2.3}) and (\ref{2.4}).
Relation (\ref{2.6}) follows from (\ref{1.8}), (\ref{2.1}),
footnote~\ref{fn2} and the multilinearity of determinant; or,
in the semi-infinite case, from (\ref{2.5}), using $\tau_0(t,s)=1$:
$$
\frac{\tau_n(t,s)}{\tau_n(-s,-t)}
= -\frac{\tau_{n-1}(t,s)}{\tau_{n-1}(-s,-t)}=\cdots
= (-1)^n\frac{\tau_0(t,s)}{\tau_0(-t,-s)}=(-1)^n.
$$
\end{proof}

For a skew-symmetric initial matrix $m_\iy(0,0)$,
relation (\ref{2.1}) implies the skew-symmetry of $m_\iy(t,-t)$.
Therefore the odd $\tau$-functions vanish and the even ones have a
natural square root, the {\em Pfaffian\/} $\TT_{2n}(t)$:
\begin{equation}
\tau_{2n+1}(t,-t)=0,\quad\tau_{2n}(t,-t)=:\TT_{2n}^2(t),
\label{2.7}
\end{equation}
where the Pfaffian, together with its sign specification, is also
determined by the formula:
\begin{equation}
\TT_{2n}(t)dx_0\wedge dx_1\wedge\dots\wedge dx_{2n-1}
:=\frac{1}{n!}\left(\sum_{0\leq i<j\leq 2n-1}
\mu_{ij}(t,-t)dx_i\wedge dx_j\right)^n.
\label{2.8}
\end{equation}

\begin{theorem}\label{thm2.2}
For $\tau$ satisfying (\ref{2.6}), and hence in particular
for a skew-symmetric initial condition $m_\iy(0,0)$, the
2-Toda $\tau$-function $\tau(t,s)$ and the
Pfaffians $\TT(t)$ are related by
\begin{align}
\begin{split}
\tau_{2n}(t+[\al]-[\beta],-t)&=
\TT_{2n}(t)\TT_{2n}(t+[\al]-[\beta]),
\\
\tau_{2n+1}(t+[\al]-[\beta],-t)&=
(\al-\beta)\TT_{2n}(t-[\beta])\TT_{2n+2}(t+[\al]),
\end{split}
\label{2.9}
\\
\intertext{or alternatively}
\begin{split}
\tau_{2n}(t-[\beta],-t+[\al])&=
\TT_{2n}(t-[\al])\TT_{2n}(t-[\beta]),
\\
\tau_{2n}(t+[\al],-t-[\beta])&=
\TT_{2n}(t+[\al])\TT_{2n}(t+[\beta]),
\\
\tau_{2n+1}(t-[\beta],-t+[\al])&=
(\al-\beta)\TT_{2n}(t-[\al]-[\beta])\TT_{2n+2}(t),
\\
\tau_{2n+1}(t+[\al],-t-[\beta])&=
(\al-\beta)\TT_{2n}(t)\TT_{2n+2}(t+[\al]+[\beta]).
\end{split}
\label{2.10}
\end{align}
\end{theorem}

\begin{proof}
In formula (\ref{1.12}), set $n=m-1$, $s=-t+[\beta]$, $t'=t+[\al]-[\beta]$ and
$s'=s-[\al]-[\beta]=-t-[\al]$; then using
\begin{multline*}
\frac{1}{2\pi i}
\oint_{z=\infty}\tau_n(t-[z^{-1}],s)\tau_{m+1}(t'+[z^{-1}],s')
e^{\sum(t_i-t_i')z^i}z^{n-m-1}dz
\\
\begin{split}
&=\frac{1}{2\pi i}\oint_{z=\infty}
\tau_{m-1}(t-[z^{-1}],s)\tau_{m+1}(t'+[z^{-1}],s')
\frac{1-\al z}{1-\beta z}\frac{dz}{z^2}
\\
&= -\Res_{z=\beta^{-1}}
\tau_{m-1}(t-[z^{-1}],s)\tau_{m+1}(t'+[z^{-1}],s')
\frac{1-\al z}{1-\beta z}\frac{dz}{z^2}
\\
&= (\beta-\al)
\tau_{m-1}(t-[\beta],s)\tau_{m+1}(t'+[\beta],s')
\\
&= (\beta-\al)
\tau_{m-1}(t-[\beta],-t+[\beta])\tau_{m+1}(t+[\al],-t-[\al]),
\end{split}
\end{multline*}
\begin{multline*}
\frac{1}{2\pi i}\oint_{z=0}\tau_m(t,s-[z])\tau_m(t',s'+[z])
e^{\sum(s_i-s_i')z^{-i}}z^{n-m-1}dz
\\
\begin{split}
&=\frac{1}{2\pi i}\oint_{z=0}
\tau_m(t,s-[z])\tau_m(t',s'+[z])
\frac{1}{1-\al/z}\frac{1}{1-\beta/z}\frac{dz}{ z^2}
\\
&=(\Res_{z=\al}+\Res_{z=\beta})\tau_m(t,s-[z])\tau_m(t',s'+[z])
\frac{dz}{(z-\al)(z-\beta)}
\\
&=\frac{1}{\al-\beta}
\bigl( \tau_m(t,s-[\al])\tau_m(t',s'+[\al]) - \tau_m(t,s-[\beta])\tau_m(t',s'+[\beta])\bigr)
\\
&=\frac{1}{\al-\beta}
\bigl( \tau_m(t,-t+[\beta]-[\al])\tau_m(t+[\al]-[\beta],-t)
\\
&\hphantom{{}=\frac{1}{\al-\beta}}\qquad
- \tau_m(t,-t)\tau_m(t+[\al]-[\beta],-t-[\al]+[\beta])\bigr),
\end{split}
\end{multline*}
and (\ref{2.6}),
we have
\begin{multline*}
-(\beta-\al)^2\tau_{m-1}(t-[\beta],-t+[\beta])\tau_{m+1}(t+[\al],-t-[\al])
\\
= (-1)^m\tau_m(t+[\al]-[\beta],-t)^2
- \tau_m(t,-t)\tau_m(t+[\al]-[\beta],-t-[\al]+[\beta]).
\end{multline*}
Setting first $m=2l$ and then $m=2l+1$, one finds respectively,
since odd $\tau$-functions vanish on $\{s=-t\}$ in view of (\ref{2.6}):
\be
0=
\tau_{2l}(t+[\al]-[\beta],-t)^2
- \tau_{2l}(t,-t)\tau_{2l}(t+[\al]-[\beta],-t-[\al]+[\beta]),
\label{2.11}
\ee
and
\begin{multline}
-(\beta-\al)^2\tau_{2l}(t-[\beta],-t+[\beta])\tau_{2l+2}(t+[\al],-t-[\al])
\\
= -\tau_{2l+1}(t+[\al]-[\beta],-t)^2.
\label{2.12}
\end{multline}
Taking the square root, with the consistent choice of sign,\footnote{
        It suffices to check that (\ref{2.8}) yields the correct sign in the
        second equation of (\ref{2.9}) at $\beta=0$, $t=0$ and
        modulo $O(\al^2)$, i.e.,
        $$
        (\pl/\pl t_1)\tau_{2n+1}(0,0)= \TT_{2n}(0)\TT_{2n+2}(0),
        $$
        for some $m_\iy(0,0)$ for which the right hand side does not vanish.
        This can be checked easily, e.g., for $m_\iy(0,0)$ made of $2\times2$
        blocks
        $\begin{pmatrix}0&1\\-1&0\end{pmatrix}$ on the diagonal.
}
(\ref{2.8}) yields (\ref{2.9}), and then (\ref{2.10}) upon setting
$t\to t-[\al]$ or $t\to t+[\beta]$.
\end{proof}

\begin{corollary}
Under the assumption of theorem~\ref{thm2.2}, the wave and adjoint wave
functions $\Psi$, $\Psi^*$ along the locus $\{s=-t\}$ satisfy the relations
\begin{align*}
\lefteqn{
\Psi_{1,2n}(t,-t,z) =
-\biggl(\frac{\tau_{2n+1}}{\sqrt{\tau_{2n}\tau_{2n+2}}}\Psi_{1,2n+1}(z)\biggr)
\biggr|_{s=-t}
}&
\\
&= \biggl(\frac{\tau_{2n+1}}{\tau_{2n}}\Psi_{2,2n}^*(z^{-1})\biggr)
\biggr|_{s=-t}
=
\Biggl(\sqrt{\frac{\tau_{2n+2}}{\tau_{2n}}}\Psi_{2,2n+1}^*(z^{-1})\Biggr)
\Biggr|_{s=-t}
\\
&=
\frac{\TT_{2n}(t-[z^{-1}])}{\TT_{2n}(t)}z^{2n}e^{\sum t_iz^i},
\\
\lefteqn{
\Psi_{1,2n-1}^*(t,-t,z) =
\biggl(\frac{\tau_{2n-1}}{\sqrt{\tau_{2n-2}\tau_{2n}}}\Psi_{1,2n-2}^*(z)\biggr)
\biggr|_{s=-t}
}&
\\
&= \biggl(\frac{\tau_{2n-1}}{\tau_{2n}}\Psi_{2,2n-1}(z^{-1})\biggr)
\biggr|_{s=-t}
= -\Biggl(\sqrt{\frac{\tau_{2n-2}}{\tau_{2n}}}\Psi_{2,2n-2}(z^{-1})\Biggr)
\Biggr|_{s=-t}
\\
&=
\frac{\TT_{2n}(t+[z^{-1}])}{\TT_{2n}(t)}z^{-(2n-1)}e^{-\sum t_iz^i}.
\end{align*}
\end{corollary}

\begin{proof}
These follow from (\ref{1.10}), (\ref{2.9}) and (\ref{2.10}) by
straightforward calculations.
\end{proof}

\begin{corollary}
Under the assumption of theorem~\ref{thm2.2}, we have\\
(i) for $k\ge1$:
\begin{align*}
\frac{\pl\tau_{2n}}{\pl t_k}\biggr|_{s=-t}
&=\TT_{2n}(t)\frac{\pl\TT_{2n}}{\pl t_k}(t),
\\
\frac{\pl\tau_{2n+1}}{\pl t_k}\biggr|_{s=-t}
&=p_{k-1}(-\tilde D_t)\TT_{2n}\cdot\TT_{2n+2}(t)
\\
&:=
\sum_{i+j=k-1}(p_i(-\tilde\pl_t)\TT_{2n}(t))(p_j(\tilde\pl_t)\TT_{2n+2}(t)),
\end{align*}
(ii) for $m\ge2$:
\begin{align*}
\sum_{k+l=m}\frac{\pl^2\tau_{2n}}{\pl t_k\pl t_l}\biggr|_{s=-t}
&=\TT_{2n}(t)\sum_{k+l=m}\frac{\pl^2\TT_{2n}}{\pl t_k\pl t_l}(t),
\\
\sum_{k+l=m}\frac{\pl^2\tau_{2n+1}}{\pl t_k\pl t_l}\biggr|_{s=-t}
&=\sum_{k+l=m-1}
 (k-l)(p_k(-\tilde\pl_t)\TT_{2n}(t))(p_l(\tilde\pl_t)\TT_{2n+2}(t)),
\\
-\sum_{k+l=m}\frac{\pl^2\tau_{2n}}{\pl t_k\pl s_l}\biggr|_{s=-t}
&=
\sum_{k+l=m}\frac{\pl\TT_{2n}}{\pl t_k}(t)\frac{\pl\TT_{2n}}{\pl s_l}(t),
\end{align*}
(iii) for $k$, $l\ge0$:
\begin{align*}
p_k(\tilde\pl_t)p_l(-\tilde\pl_t)\tau_{2n}(t,s)|_{s=-t}
&=
\tilde\tau_{2n}(t)p_k(\tilde\pl_t)p_l(-\tilde\pl_t)\tilde\tau_{2n}(t),
\\
p_k(\tilde\pl_t)p_l(-\tilde\pl_t)\tau_{2n+1}(t,s)|_{s=-t}
&=p_l(-\tilde\pl_t)\tilde\tau_{2n}(t)\cdot
 p_{k-1}(\tilde\pl_t)\tilde\tau_{2n+2}(t)
 \\
&\phantom{=}-p_{l-1}(-\tilde\pl_t)\tilde\tau_{2n}(t)\cdot
 p_k(\tilde\pl_t)\tilde\tau_{2n+2}(t),
\end{align*}
where $p_k(\cdot)$ are the elementary Schur functions, with $p_{-1}(\cdot)=0$,
and $D_t=(D_{t_1},(1/2)D_{t_2},\dots)$ are Hirota's symbols.
\end{corollary}

\begin{proof} Relations (i) are obtained by differentiating formulas
(\ref{2.9}) in $\al$, setting $\beta=\al$ and identifying the coefficients of
$\al^{k-1}$.  The first two relations in (ii) are obtained by
differentiating formulas
(\ref{2.9}) in $\al$ and $\beta$ (i.e., applying $\pl^2/\pl\al\pl\beta$),
setting $\beta=\al$ and identifying the coefficients of $\al^{m-2}$.
The last relation in (ii) is obtained by differentiating the first
formula in (\ref{2.10}) in $\al$ and $\beta$, setting $\beta=\al$,
substituting $t+[\al]$ for $t$, and then identifying the coefficients
of $\al^{m-2}$.
Finally, expanding both identities (\ref{2.9}) in $\al$ and $\beta$, e.g.,
$$
 \tau_{2n}(t+[\al]-[\beta],s)=\sum_{k,l=0}^\iy\al^k\beta^l
 p_k(\tilde\pl_t)p_l(-\tilde\pl_t)\tau_{2n}(t,s)
$$
and identifying the powers of $\al$ and $\beta$ yields relations (iii).
\end{proof}

Variants of formulas (\ref{2.9}) and the formulas in the corollary
can be obtained by using (\ref{2.6}) and the following consequence of it:
$$
\frac{\pl^{|I|+|J|}}{\pl t^I\pl s^J}\tau_n\biggr|_{s=-t}
=(-1)^{|I|+|J|+n}\frac{\pl^{|J|+|I|}}{\pl t^J\pl s^I}\tau_n\biggr|_{s=-t},
$$
where $I=(i_1,i_2,\dots)$ and $J=(j_1,j_2,\dots)$ are multiindices,
$|I|=i_1+i_2+\cdots$, $\pl t^I=\pl t_1^{i_1}\pl t_2^{i_2}\cdots$ etc.
In particular, $(\pl^2/\pl t_k\pl s_l+\pl^2/\pl t_l\pl s_k)\tau_{2n+1}=0$,
so we get the (rather trivial) counterpart of the last formula in part (ii)
of the corollary:
$$
\sum_{k+l=m}\frac{\pl^2\tau_{2n+1}}{\pl t_k\pl s_l}\biggr|_{s=-t} = 0.
$$

\section{Equations satisfied by Pfaffian $\TT$-functions}

In this section, we exhibit the properties of the Pfaffian $\TT$-function
introduced above, for the 2-Toda $\tau$-function satisfying (\ref{2.6}),
or the skew-symmetric initial data $m_\iy(0,0)$.
As in the last section, whenever we make a connection with the matrix
$m_\iy$, we assume either $m_\iy$ is semi-infinite, or
$\det m_n$ can be interpreted as in formula~$(*)$ in footnote~\ref{fn2}.

\begin{theorem}
The $\TT$-functions satisfy the bilinear relations
\begin{multline}
\oint_{z=\iy}\TT_{2n}(t-[z^{-1}])\TT_{2m+2}(t'+[z^{-1}])
e^{\sum_0^\iy(t_i-t'_i)z^i}
z^{2n-2m-2}dz
\\
{}+\oint_{z=0}\TT_{2n+2}(t+[z])\TT_{2m}(t'-[z])
e^{\sum_0^\iy(t'_i-t_i)z^{-i}}z^{2n-2m}dz=0,
\label{3.1}
\end{multline}
or equivalently\footnote{
        $\tilde\pl=(
        \frac{\pl}{\pl t_1},\frac{1}{2}\frac{\pl}{\pl t_2},
        \frac{1}{3}\frac{\pl}{\pl t_3},\dots)$, and
        $\tilde D=\left(
        D_1,\frac{1}{2}D_2,\frac{1}{3}D_3,\dots\right)$ is the corresponding
        Hirota symbol, i.e.,
        $P(\tilde D)f\cdot g
        := P(\pl/\pl y_1,\frac{1}{2}\pl/\pl y_2,\frac{1}{3}\pl/\pl y_3,\dots)
        f(t+y)g(t-y)|_{y=0}$ for any polynomial $P$; and $p_k$ are the
        elementary Schur functions:
        $\sum_0^\infty p_k(t)z^k:=\exp(\sum_1^\infty t_iz^i)$.
}
\begin{multline}
\sum_{\substack{j,k\ge0\\j-k=-2n+2m+1}}
p_j(-2y) e^{\sum-y_iD_i}p_k(-\tilde D)\TT_{2n}\cdot\TT_{2m+2}
\\
+ \sum_{\substack{j,k\ge0\\k-j=-2n+2m-1}}
p_j(2y) e^{\sum-y_iD_i}p_k(\tilde D)\TT_{2n+2}\cdot\TT_{2m}
=0.
\label{3.2}
\end{multline}
\end{theorem}

\begin{proof}
Formula (\ref{3.1}) follows from (\ref{1.12}) upon replacing $n$ by $2n$
and $m$ by $2m$,\footnote{
        One can check that all the other choices of parities of $n$ and $m$
        yield the same formula.
}
using (\ref{2.9}) and (\ref{2.10}), with $\beta=0$,
to eliminate $\tau_{2n}(t-[z^{-1}],-t)$, $\tau_{2m+1}(t',-t'-[z])$,
$\tau_{2n+1}(t,-t-[z])$ and $\tau_{2m}(t'-[z],-t')$ and, upon dividing
both sides by $\TT_{2n}(t)\TT_{2m}(t)$.

Substituting $t-y$ and $t+y$ for $t$ and $t'$, respectively, into
the left hand side of (\ref{3.1}) and Taylor expanding it in $y$, we obtain
\begin{multline*}
\oint_{z=\infty}
e^{-\sum_1^\infty2y_iz^i} \TT_{2n}(t-y-[z^{-1}])\TT_{2m+2}(t+y+[z^{-1}])
z^{2n-2m-2}dz
\\
{}+ \oint_{z=0}
e^{\sum_1^\infty2y_iz^{-i}} \TT_{2n+2}(t-y+[z])\TT_{2m}(t+y-[z])
z^{2n-2m}dz
\\
\begin{split}
&= \oint_{z=\infty}
e^{-\sum_1^\infty2y_iz^i} e^{\sum-y_iD_i}e^{-\sum z^{-i}D_i/i}
\TT_{2n}\cdot\TT_{2m+2}
z^{2n-2m-2}dz
\\
&\phantom{={}}+ \oint_{z=0}
e^{\sum_1^\infty2y_iz^{-i}} e^{\sum-y_iD_i}e^{\sum z^iD_i/i}
\TT_{2n+2}\cdot\TT_{2m}
z^{2n-2m}dz
\\
&= \oint_{z=\infty}
\sum_{j=0}^\infty p_j(-2y)z^j
e^{\sum-y_iD_i}\sum_{k=0}^\infty p_k(-\tilde D)z^{-k}
\TT_{2n}\cdot\TT_{2m+2}
z^{2n-2m-2}dz
\\
&\phantom{={}}+\oint_{z=0}
\sum_{j=0}^\infty p_j(2y)z^{-j}
e^{\sum-y_iD_i}\sum_{k=0}^\infty p_k(\tilde D)z^k
\TT_{2n+2}\cdot\TT_{2m}
z^{2n-2m}dz
\\
&= 2\pi i\biggl(\sum_{j-k=-2n+2m+1}
p_j(-2y) e^{\sum-y_iD_i}p_k(-\tilde D) \TT_{2n}\cdot\TT_{2m+2}
\\
&\hphantom{=2\pi i\biggl(}+ \sum_{k-j=-2n+2m-1}
p_j(2y) e^{\sum-y_iD_i}p_k(\tilde D) \TT_{2n+2}\cdot\TT_{2m}\biggr),
\end{split}
\end{multline*}
showing the equivalence of (\ref{3.1}) and (\ref{3.2}).
\end{proof}

The identity (\ref{3.1}) gives various bilinear relations satisfied
by $\TT$.  We show that the Pfaffian $\TT$-functions satisfy
identities reminiscent of the Fay and differential Fay identities
for the KP or 2-Toda $\tau$-functions (e.g., see \cite{AvM1}). From
this we deduce a sequence of Hirota bilinear equations for $\TT$,
which can be interpreted as a recursion relation for $\TT_{2n}(t)$.

\begin{theorem}
The functions $\TT_{2n}(t)$ satisfy the following ``Fay identity'':
\begin{multline}
\sum_{i=1}^r
\TT_{2n}\biggl(t-\sum_{j=1}^l[z_j]-[\zeta_i]\biggr)
\TT_{2m+2}\biggl(t-\sum_{\substack{1\le j\le r\\j\ne i}}[\zeta_j]\biggr)
\frac{\prod_{k=1}^l(\zeta_i-z_k)}
{\prod_{\substack{1\le k\le r\\k\ne i}}(\zeta_i-\zeta_k)}
\\
+ \sum_{i=1}^l
\TT_{2n+2}\biggl(t-\sum_{\substack{1\le j\le l\\j\ne i}}[z_j]\biggr)
\TT_{2m}\biggl(t-\sum_{j=1}^r[\zeta_j]-[z_i]\biggr)
\frac{\prod_{k=1}^r(z_i-\zeta_k)}
{\prod_{\substack{1\le k\le l\\k\ne i}}(z_i-z_k)}=0,
\label{Fay}
\end{multline}
the ``differential Fay identity'':
\begin{multline}
\{\TT_{2n}(t-[u]),\TT_{2n}(t-[v])\}
\\
{}+(u^{-1}-v^{-1})(\TT_{2n}(t-[u])\TT_{2n}(t-[v])
-\TT_{2n}(t)\TT_{2n}(t-[u]-[v]))
\\
=uv(u-v)\TT_{2n-2}(t-[u]-[v])\TT_{2n+2}(t),
\label{3.3}
\end{multline}
and Hirota bilinear equations, involving nearest neighbors:
\begin{equation}
\left(p_{k+4}(\tilde D)-\frac{1}{2}D_1D_{k+3}
\right)\TT_{2n}\cdot\TT_{2n}=p_k(\tilde D)\TT_{2n+2}\cdot\TT_{2n-2}.
\label{3.4}
\end{equation}
Here, in (\ref{Fay}) $2n$, $2m\in A$, $l$, $r\ge0$ such that $r-l=2n-2m$, $z_i$
($1\le i\le l$) and $\zeta_i$ ($1\le i\le r$) are scalar parameters near $0$;
in (\ref{3.3}) $2n-2\in A$ (hence $2n$, $2n+2\in A$), and $u$ and $v$ are
scalar parameters near $0$; and in (\ref{3.4}) $2n-2\in A$,
$k=0$, $1$, $2$, \dots,
and $\{f,g\}:=f'g-fg'=D_1f\cdot g$ is the Wronskian of $f$ and $g$,
where $'=\pl/\pl t_1$.
\end{theorem}

\begin{proof}
The Fay identity (\ref{Fay}) follows from the bilinear identity (\ref{3.1})
by substitutions
$$
t\mapsto t-[z_1]-\cdots-[z_l]\quad\text{and}\quad
t'\mapsto t-[\zeta_1]-\dots-[\zeta_r].
$$
Indeed, since $r-l=2n-2m$, we have
\begin{align*}
\exp\bigl(\sum(t_i-t_i')z^i\bigr)z^{2n-2m-2}dz
&=
\frac{\prod_{k=1}^l(1-zz_k)}{\prod_{k=1}^r(1-z\zeta_k)}z^{r-l-2}dz
\\
&=
-\frac{\prod_{k=1}^l((1/z)-z_k)}{\prod_{k=1}^r((1/z)-\zeta_k)}d(1/z)
\end{align*}
and
\begin{align*}
\exp\bigl(\sum(t_i'-t_i)z^{-i}\bigr)z^{2n-2m}dz
&=
\frac{\prod_{k=1}^r(1-\zeta_k/z)}{\prod_{k=1}^l(1-z_k/z)}z^{r-l}dz
\\
&=
\frac{\prod_{k=1}^r(z-\zeta_k)}{\prod_{k=1}^l(z-z_k)}dz,
\end{align*}
so the first and second terms on the left hand side of (\ref{3.1}),
divided by $2\pi i$, become, respectively,
\begin{align*}
\lefteqn{
\frac{1}{2\pi i}\oint_{z=\iy}\dots
}&
\\
&=-\sum_{i=1}^r\Res_{z=\zeta_i^{-1}}
\TT_{2n}\biggl(t-\sum_{j=1}^l[z_j]-[z^{-1}]\biggr)
\\
&\qquad\qquad\qquad\cdot
\TT_{2m+2}\biggl(t-\sum_{j=1}^r[\zeta_j]+[z^{-1}]\biggr)
\frac{\prod_{k=1}^l(1-zz_k)}{\prod_{k=1}^r(1-z\zeta_k)}dz
\displaybreak[0]\\
&=\sum_{i=1}^r\Res_{\zeta=\zeta_i}
\TT_{2n}\biggl(t-\sum_{j=1}^l[z_j]-[\zeta]\biggr)
\\
&\qquad\qquad\qquad\cdot
\TT_{2m+2}\biggl(t-\sum_{j=1}^r[\zeta_j]+[\zeta]\biggr)
\frac{\prod_{k=1}^l(\zeta-z_k)}{\prod_{k=1}^r(\zeta-\zeta_k)}d\zeta
\qquad(\zeta:=z^{-1})
\displaybreak[0]\\
&=\sum_{i=1}^r
\TT_{2n}\biggl(t-\sum_{j=1}^l[z_j]-[\zeta_i]\biggr)
\TT_{2m+2}\biggl(t-\sum_{j=1}^r[\zeta_j]+[\zeta_i]\biggr)
\frac{\prod_{k=1}^l(\zeta_i-z_k)}
{\prod_{\substack{1\le k\le r\\k\ne i}}(\zeta_i-\zeta_k)},
\end{align*}
and
\begin{align*}
\lefteqn{
\frac{1}{2\pi i}\oint_{z=0}\dots
}&
\\
&=\sum_{i=1}^l\Res_{z=z_i}
\TT_{2n+2}\biggl(t-\sum_{j=1}^l[z_j]+[z]\biggr)
\\
&\qquad\qquad\qquad\cdot
\TT_{2m}\biggl(t-\sum_{j=1}^r[\zeta_j]-[z]\biggr)
\frac{\prod_{k=1}^r(z-\zeta_k)}{\prod_{k=1}^l(z-z_k)}dz
\displaybreak[0]\\
&=\sum_{i=1}^l
\TT_{2n+2}\biggl(t-\sum_{j=1}^l[z_j]+[z_i]\biggr)
\TT_{2m}\biggl(t-\sum_{j=1}^r[\zeta_j]-[z_i]\biggr)
\frac{\prod_{k=1}^r(z_i-\zeta_k)}
{\prod_{\substack{1\le k\le l\\k\ne i}}(z_i-z_k)},
\end{align*}
showing (\ref{Fay}).

Note that when $2m=2n-2$, $l=1$ and $r=3$, denoting $z_i=\zeta_{i-1}$ for
$2\le i\le4$, and multiplying both sides of (\ref{Fay}) by
$\prod_{2\le j<k\le4}(z_j-z_k)$, we obtain
\begin{multline}
(z_2-z_1)(z_3-z_4)\TT_{2n}(t-[z_1]-[z_2])\TT_{2n}(t-[z_3]-[z_4])
\\
-(z_3-z_1)(z_2-z_4)\TT_{2n}(t-[z_1]-[z_3])\TT_{2n}(t-[z_2]-[z_4])
\\
+(z_4-z_1)(z_2-z_3)\TT_{2n}(t-[z_1]-[z_4])\TT_{2n}(t-[z_2]-[z_3])
\\
+\biggl(\prod_{1\le i<j\le4}(z_i-z_j)\biggr)
\TT_{2n+2}(t)\TT_{2n-2}(t-[z_1]-[z_2]-[z_3]-[z_4])=0.
\label{basicFay}
\end{multline}
The differential Fay identity (\ref{3.3}) follows from (\ref{basicFay})
by taking a limit (set $z_4=0$, divide by $z_3$ and let $z_3\to0$).
Alternatively, we can prove (\ref{3.3})
directly from (\ref{3.1}): Set $t-t'=[u]-[v]$, $2m=2n-2$ in (\ref{3.1})
and in the clockwise integral about $z=\iy$, set $z\mapsto 1/z$, yielding
\begin{multline*}
\oint_0\TT_{2n}(t-[z])\TT_{2n}(t'+[z])\frac{1-v/z}{1-u/z}\,
\frac{dz}{z^2}
\\
=-\oint_0\TT_{2n+2}(t+[z])\TT_{2n-2}(t'-[z])\frac{1-u/z}{1-v/z}
z^2dz.
\end{multline*}
The first integral has a simple pole at $z=u$ and a double pole at
$z=0$, while the second integral has a simple pole at $z=v$ only,
yielding, after substitution $t'=t-[u]+[v]$,
\begin{multline*}
\TT_{2n}(t-[u])\TT_{2n}(t+[v])(u-v)\frac{1}{u^2}
\\
+\frac{d}{dz}\biggl(
\TT_{2n}(t-[z])\TT_{2n}(t-[u]+[v]+[z])\frac{z-v}{z-u}\biggr)\biggr|_{z=0}
\\
=-\TT_{2n+2}(t+[v])\TT_{2n-2}(t-[u])(v-u)v^2,
\end{multline*}
or, after carrying out $d/dz|_{z=0}$ on the left hand side,
\begin{multline}
\TT_{2n}(t-[u])\TT_{2n}(t+[v])(u-v)\frac{1}{u^2}
\\
+\TT_{2n}(t)\cdot\TT_{2n}(t-[u]+[v])\frac{v-u}{u^2}
- D_1\,\TT_{2n}(t)\cdot\TT_{2n}(t-[u]+[v])\frac{v}{u}
\\
=-\TT_{2n+2}(t+[v])\TT_{2n-2}(t-[u])(v-u)v^2.
\label{3.5}
\end{multline}
Shifting $t\mapsto t-[v]$ and multiplying both sides by $u/v$ yield
(\ref{3.3}).

Since $P(-D)f\cdot f=P(D)f\cdot f$ by the definition of Hirota operator,
(\ref{3.4}) is the same as (\ref{0.10}), which, as we have pointed out, are
nothing but the coefficients of $y_{k+3}$ in (\ref{0.8}), or (\ref{3.2}).
It also follows from (\ref{3.3}), since, for any power series $F(t,t')$
which satisfies $F(t,t)\equiv0$,
\begin{multline*}
\text{coefficient of $y_{k+3}$ in $F(t-y,t+y)$}
=\biggl(\frac{\pl}{\pl t_n'}-\frac{\pl}{\pl t_n}\biggr)F(t,t)
\\
=2\frac{\pl}{\pl t_n'}F(t,t')
=2\times\text{coefficient of $u^{k+2}$ in }
\frac{d}{dv}F(t,t-[u]+[v])\biggr|_{v=u}.
\end{multline*}
Indeed, differentiating  (\ref{3.5}), which is equivalent
to (\ref{3.3}), in $v$, setting $v=u$ and using $D_1f\cdot f=0$,
\begin{align*}
\frac{\pl}{\pl v}(D_1f(t)\cdot g(t+[v]))
&=-\frac{1}{2}D_1D_vf(t)\cdot g(t+[v])
\\
&=-\frac{1}{2}\sum v^{j-1}D_1D_jf(t)\cdot g(t+[v]),
\end{align*}
etc., we have
\begin{gather*}
-\TT_{2n}(t-[u])\TT_{2n}(t+[u])\frac{1}{ u^2}
+\TT_{2n}(t)^2\frac{1}{ u^2}
+ \frac{1}{2}\sum_{j=1}^\iy u^{j-1}D_1D_j\TT_{2n}\cdot\TT_{2n}(t)
\\
=-\TT_{2n+2}(t+[u])\TT_{2n-2}(t-[u])u^2,
\end{gather*}
which, noting $f(t+[u])g(t-[u])=\sum u^kp_k(\tilde D)f\cdot g$, is
a generating function for (\ref{3.4}).
\end{proof}

As in the case of KP or 2-Toda $\tau$-functions, Pfaffian $\TT$-functions
satisfy higher degree identities:

\begin{theorem}
\begin{multline}
\Pf\biggl(
\frac{(z_j-z_i)\TT_{2n-2}(t-[z_i]-[z_j])}{\TT_{2n}(t)}
\biggr)_{1\le i,j\le 2k}
\\
=\Dt(z)\frac{\TT_{2n-2k}\Bigl(t-\sum_1^{2k}[z_i]\Bigr)}{\TT_{2n}(t)},
\label{PfaffFay}
\end{multline}
where $k\ge1$, $2n-2k\in A$, $z_i$ are scalar parameters near $0$, and
$\Dt(u_1,\dots,u_n):=\prod_{i<j}(u_j-u_i)$.
\end{theorem}

\begin{proof}
This may be obtained, up to the sign, from the second identity in
Theorem~4.2 of \cite{AvM3}:
\begin{multline*}
\det\biggl(
\frac{\tau_{N-1}(t-[z_i],s+[y_j])}{\tau_N(t,s)}
\biggr)_{1\le i,j\le k}
\\
=\Dt(y)\Dt(z)
\frac{\tau_{N-k}\Bigl(t-\sum_1^k[z_i],s+\sum_1^k[y_i]\Bigr)}{\tau_N(t,s)},
\end{multline*}
by setting $N\mapsto2n$, $k\mapsto2k$, $y_i=z_i$, taking the square roots
of both sides and using (\ref{2.10}).
Rather than taking this route, here we prove (\ref{PfaffFay}) by induction
on $k$, using the bilinear Fay identity (\ref{Fay}).  First, (\ref{PfaffFay})
is trivial when $k=1$. (Note also that it gives (\ref{basicFay}) when $k=2$.)
Suppose (\ref{PfaffFay}) holds for $k-1$.  Then we have, for every
$p\in\{2,\dots,2k\}$,
\begin{multline*}
\Pf\biggl(
\frac{(z_j-z_i)\TT_{2n-2}(t-[z_i]-[z_j])}{\TT_{2n}(t)}
\biggr)_{\substack{2\le i,j\le2k\\i,j\ne p}}
\\
=\Dt(z_2,\dots,\widehat{z_p},\dots,z_{2k})
\frac{\TT_{2n-2k+2}\Bigl(t-\sum_{2\le i\le2k,i\ne p}[z_i]\Bigr)}{\TT_{2n}(t)}.
\end{multline*}
Multiplying both sides by
$(-1)^p(z_p-z_1)\TT_{2n-2}(t-[z_1]-[z_p])/\TT_{2n}(t)$,
summing it up for $p=2,\dots,2k$, and using
$$
(-1)^p(z_p-z_1)\Dt(z_2,\dots,\widehat{z_p},\dots,z_{2k})
=\frac{\Dt(z)}{\prod_{2\le i\le2k}(z_i-z_1)}
\frac{(z_p-z_1)}{\prod_{\substack{2\le i\le2k\\i\ne p}}(z_p-z_i)}
$$
and the identity
$$
\Pf(a_{ij})_{1\le i,j\le2k}=\sum_{p=2}^{2k}(-1)^p
a_{1p}\Pf(a_{ij})_{\substack{2\le i,j\le2k\\i,j\ne p}}
\qquad \forall(a_{ij})\text{ skew symmetric}
$$
which follows from definition (\ref{2.8}) of the Pfaffian, we have
\begin{align*}
\lefteqn{
\Pf\biggl(
\frac{(z_j-z_i)\TT_{2n-2}(t-[z_i]-[z_j])}{\TT_{2n}(t)}
\biggr)_{1\le i,j\le 2k}
}\qquad&
\\
&=\frac{\Dt(z)}{\prod_{2\le i\le2k}(z_i-z_1)}\cdot\frac{1}{\TT_{2n}(t)^2}
\sum_{p=2}^{2k}
\frac{(z_p-z_1)}{\prod_{\substack{2\le i\le2k\\i\ne p}}(z_p-z_i)}
\\
&\qquad\qquad{}\cdot
\TT_{2n-2}(t-[z_1]-[z_p])
\TT_{2n-2k+2}\biggl(t-\sum_{\substack{2\le i\le2k\\i\ne p}}[z_i]\biggr)
\\
\intertext{%
using the bilinear Fay identity (\ref{Fay}) with $r=2k-1$, $l=1$,
$\zeta_i:=z_{i+1}$ ($1\le i\le2k-1$) and $(2n,2m)$ replaced by
$(2n-2,2n-2k)$ this becomes}
&=\frac{\Dt(z)}{\prod_{2\le i\le2k}(z_i-z_1)}\cdot\frac{1}{\TT_{2n}(t)^2}
\\
&\qquad\qquad{}\cdot
(-1)\biggl(\prod_{i=2}^{2k}(z_1-z_i)\biggr)
\TT_{2n}(t)\TT_{2n-2k}\biggl(t-\sum_{i=1}^{2k}[z_i]\biggr)
\\
&=\Dt(z)\frac{\TT_{2n-2k}\Bigl(t-\sum_{i=1}^{2k}[z_i]\Bigr)}{\TT_{2n}(t)},
\end{align*}
completing the proof of (\ref{PfaffFay}) by induction.
\end{proof}

\section{Vertex operators for Pfaffian $\TT$-functions}

In terms of the operators
\begin{align*}
X(t,\lb)&:=e^{\sum_1^\iy t_k\lb^k}
e^{-\sum_1^\iy\frac{\lb^{-k}}{k}\frac{\pl}{\pl t_k}},
\\
X^*(t,\lb)&:=e^{-\sum_1^\iy t_k\lb^k}
e^{\sum_1^\iy\frac{\lb^{-k}}{k}\frac{\pl}{\pl t_k}}
\end{align*}
acting on functions $f(t)$ of $t=(t_1,t_2,\dots)\in\BC^\iy$, define
the following four operators\footnote{
        \label{afn1}
        Here $X(s,\lb)$ has $s_i$ in place of $t_i$, as well as
        $\pl/\pl s_i$ in place of $\pl/\pl t_i$, in the definition
        of $X(t,\lb)$, etc.; $\chi(\mu):=(\mu^n)_{n\in A}$, and
        $\chi^*(\mu)=\chi(\mu^{-1})$.
}
acting on column vectors $g=(g_n(t))_{n\in A}$,
\begin{alignat*}{2}
\BX_1(\mu)      
&:=X(t,\mu)\chi(\mu),&\quad
\BX_1^*(\lb)    
&:=-\chi^*(\lb)X^*(t,\lb),
\\
\BX_2(\mu)      
&:=-X(s,\mu)\chi^*(\mu)\Lb,&\quad
\BX_2^*(\lb)    
&:=\Lb^\top\chi(\lb)X^*(s,\lb),
\end{alignat*}
and their compositions\footnote{
        \label{fn8}
        When $i=j$, $\BX_j^*$ interacts with $\BX_i$ nontrivially,
        yielding the factor $\exp(\sum(\mu/\lb)^k/k)=1/(1-\mu/\lb)$
        if we bring the multiplication operators to the left and the
        shift operators in $t$ or $s$ to the right.  So if we denote
        by :~: the usual normal ordering of operators in $t$, $s$
        (but not in the discrete index $n$), we have
        $$
        \BX_{ii}(\mu,\lb)=
        1/(1-\mu/\lb))\,\mathopen:\BX_i^*(\lb)\BX_i(\mu)\mathclose:
        = - 1/(1-\mu/\lb)\phi(\mu,\lb)X(u,\mu,\lb),
        $$
        where $u=t$, $\phi(\mu,\lb)=\chi(\mu/\lb)$ if $i=1$;
        $u=s$, $\phi(\mu,\lb)=\Lb^\top\chi(\lb/\mu)\Lb$ if $i=2$; and
        $$
        X(u,\mu,\lb):=e^{\sum_1^\iy u_k(\mu^k-\lb^k)}
        e^{\sum_1^\iy\frac{\lb^{-k}-\mu^{-k}}{k}\frac{\pl}{\pl u_k}}.
        $$
}
$$
\BX_{ij}(\mu,\lb):=\BX_j^*(\lb)\BX_i(\mu),\quad i,j=1,2.
$$
They form a set of four vertex operators associated with the 2-Toda
lattice.  Among those, $\BX_{12}$ is important in the semi-infinite
case, related to the study of orthogonal polynomials. In
\cite{AvM3}, we showed that
\begin{equation}
\sum_{m\le j<n}\Psi_{1,j}(\mu)\Psi_{2,j}^*(\lb^{-1})
=\frac{(\BX_{12}(\mu,\lb)\tau)_n}{\tau_n}
-\frac{(\BX_{12}(\mu,\lb)\tau)_m}{\tau_m}
\label{4.1}
\end{equation}
for any $n$, $m\in A$, $n\ge m$. Note on the right hand side the limit
exists as $s\to-t$ if $n$ and $m$ are even, so in particular, taking
$n=m+1$, we see the poles along $s=-t$ cancel out in
$\Psi_{1,2m}(\mu)\Psi_{2,2m}^*(\lb^{-1})
+\Psi_{1,2m+1}(\mu)\Psi_{2,2m+1}^*(\lb^{-1})$.  We shall come back to
this point after proving the following theorem and its corollary.

Suppose $\tau$ satisfies (\ref{2.6}), and let $\TT$ be the vector of
corresponding Pfaffian $\TT$-functions. Let $\BX_1$, $\BX_1^*$ and $\BX_{11}$
act on $\TT$ as if they are acting on the extended vector $(\TT_n)_{n\in A}$,
where $\TT_n\equiv0$ if $n$ is odd, so that $\chi(\mu)$ (resp.\ $\chi^*(\lb)$)
acts on $\TT_{2n}$ by multiplication of $\mu^{2n}$ (resp.\ $\lb^{-2n}$).
Then we have\footnote{
        The product $\BX_1(\lb)\BX_1(\mu)$ in (\ref{4.4})
        is computed the same way as in footnote~\ref{fn8}.
}
\begin{theorem}
\begin{align}
(\BX_{11}(\mu,\lb)\tau)_N|_{s=-t}&=
\begin{cases}
        \TT_{2n}(t)\BX_{11}(\mu,\lb)\TT_{2n}(t) &(N=2n)
        \\
        -\lb(\BX_1(\mu)\TT_{2n}(t))\BX_1^*(\lb)\TT_{2n+2}&(N=2n+1)
\end{cases}
\label{4.2}
\\
(\BX_{22}(\mu,\lb)\tau)_N|_{s=-t}&=
\begin{cases}
        -\TT_{2n}(t)\BX_{11}(\lb,\mu)\TT_{2n}(t) &(N=2n)
        \\
        -\mu(\BX_1(\lb)\TT_{2n})(\BX_1^*(\mu)\TT_{2n+2})&(N=2n+1)
\end{cases}
\label{4.3}
\\
(\BX_{12}(\mu,\lb)\tau)_N|_{s=-t}&=
\begin{cases}
        -\lb\TT_{2n}(t)\BX_1(\lb)\BX_1(\mu)\TT_{2n-2}(t) &(N=2n)
        \\
        (\BX_1(\mu)\TT_{2n})(\BX_1^*(\lb)\TT_{2n})&(N=2n+1)
\end{cases}
\label{4.4}
\end{align}
\end{theorem}

\begin{corollary}
For $k=1$, $2$, the following holds:
\begin{align*}
J_i^{(k)}(t)\tau_{2n}(t,s)|_{s=-t}&=\TT_{2n}(t)J_i^{(k)}(t)\TT_{2n}(t),
\\
J_i^{(k)}(s)\tau_{2n}(t,s)|_{s=-t}&=(-1)^k\TT_{2n}(t)J_i^{(k)}(t)\TT_{2n}(t),
\end{align*}
and so
$$
(J_i^{(k)}(t)+(-1)^kJ_i^{(k)}(s))\tau_{2n}(t,s)|_{s=-t}
=2\TT_{2n}(t)J_i^{(k)}(t)\TT_{2n}(t).
$$
\end{corollary}

\begin{proof}
The theorem follows from formulas (\ref{2.9}) and (\ref{2.10})
by straightforward calculations:
\begin{align*}
\lefteqn{(\BX_{11}(\mu,\lb)\tau)_N|_{s=-t}}\qquad&
\\
&=-\left(\frac{\mu}{\lb}\right)^N\frac{\lb}{\lb-\mu}
e^{\sum t_i(\mu^i-\lb^i)}\tau_N(t-[\mu^{-1}]-[\lb^{-1}],-t)
\\
\intertext{for $N=2n$:}
&=-\left(\frac{\mu}{\lb}\right)^N\frac{\lb}{\lb-\mu}
        e^{\sum t_i(\mu^i-\lb^i)}\TT_{2n}(t)\TT_{2n}(t-[\mu^{-1}]+[\lb^{-1}])
        \\
&=\TT_{2n}(t)\BX_{11}(\mu,\lb)\TT_{2n}(t),
\\
\intertext{for $N=2n+1$:}
&=\left(\frac{\mu}{\lb}\right)^N
        \frac{\lb(\lb^{-1}-\mu^{-1})}{\lb-\mu}
        e^{\sum t_i(\mu^i-\lb^i)}
        \TT_{2n}(t-[\mu^{-1}])\TT_{2n+2}(t+[\lb^{-1}])
        \\
        &=-\lb(\BX_1(\mu)\TT_{2n})(\BX_1^*(\lb)\TT_{2n+2});
\displaybreak[1]\\[10pt]
\lefteqn{(\BX_{22}(\mu,\lb)\tau)_N|_{s=-t}}\qquad&
\\
&=-\left(\frac{\lb}{\mu}\right)^{N-1}\frac{\lb}{\lb-\mu}
e^{-\sum t_i(\mu^i-\lb^i)}\tau_N(t,-t-[\mu^{-1}]+[\lb^{-1}])
\\
&=-\left(\frac{\lb}{\mu}\right)^{N-1}\frac{\lb}{\lb-\mu}
e^{-\sum t_i(\mu^i-\lb^i)}(-1)^N\tau_N(t+[\mu^{-1}]-[\lb^{-1}],-t)
\\
\intertext{for $N=2n$:}
&=-\left(\frac{\lb}{\mu}\right)^{2n-1}\frac{\lb}{\lb-\mu}
        e^{-\sum t_i(\mu^i-\lb^i)}\TT_{2n}(t)\TT_{2n}(t+[\mu^{-1}]-[\lb^{-1}])
        \\
&=-\TT_{2n}(t)\BX_{11}(\lb,\mu)\TT_{2n}(t),
\\
\intertext{for $N=2n+1$:}
&=\left(\frac{\lb}{\mu}\right)^{2n}
        \frac{\lb(\mu^{-1}-\lb^{-1})}{\lb-\mu}
        e^{-\sum t_i(\mu^i-\lb^i)}
        \TT_{2n}(t-[\lb^{-1}])\TT_{2n+2}(t+[\mu^{-1}])
        \\
&=\frac{\lb^{2n}}{\mu^{2n+1}}
        e^{-\sum t_i(\mu^i-\lb^i)}
        \TT_{2n}(t-[\lb^{-1}])\TT_{2n+2}(t+[\mu^{-1}])
        \\
&=-\mu(\BX_1(t,\lb)\TT_{2n})(\BX_1(t,\mu)\TT_{2n+2});
\displaybreak[1]\\[10pt]
\lefteqn{(\BX_{12}(\mu,\lb)\tau)_N|_{s=-t}}\qquad&
\\
&=(\Lb^\top\chi(\mu\lb)X^*(s,\lb)X(t,\mu)\tau)_N|_{s=-t}
\\
&=(\mu\lb)^{N-1}e^{\sum(t_i\mu^i-(-t_i)\lb^i)}
\tau_{N-1}(t-[\mu^{-1}],-t+[\lb^{-1}])
\\
\intertext{for $N=2n$:}
&=(\mu\lb)^{2n-1}e^{\sum t_i(\mu^i+\lb^i)}
        (\lb^{-1}-\mu^{-1})\TT_{2n-2}(t-[\lb^{-1}]-[\mu^{-1}])\TT_{2n}(t)
        \\
&=-\lb(\BX_1(\lb)\BX_1(\mu)\TT_{2n-2}(t))\TT_{2n}(t),
\\
\intertext{for $N=2n+1$:}
&=(\mu\lb)^{2n}e^{\sum t_i(\mu^i+\lb^i)}
        \TT_{2n}(t-[\mu^{-1}])\TT_{2n}(t-[\lb^{-1}])
        \\
&=(\BX_1(\mu)\TT_{2n})(\BX_1(\lb)\TT_{2n}).
\end{align*}
The corollary is shown by expanding $\BX_{11}$ in $\lb$ and $\mu-\lb$.
Recall that
\begin{align*}
\BX_{11}(\mu,\lb)&=-\frac{\lb}{\lb-\mu}\biggl(\left(\frac{\mu}{\lb}\right)^n
X(\mu,\lb)\biggr)_{n\in A}
\\
&=-\frac{\lb}{\lb-\mu}\biggl(\sum_{k=0}^\iy\frac{(\mu-\lb)^k}{k!}
\sum_{l=-\iy}^\iy
\lb^{-l-k}W_{n,l}^{(k)}(t)\biggr)_{n\in A},
\end{align*}
where $X(\mu,\lb)$ is the vertex operator in the KP theory
\cite{Date}, and
$$
W_{n,l}^{(k)}(t)=\sum_{j=0}^k\binom nj (k)_jW_l^{(k-j)},
$$
with $W_l^{(k)}$ the coefficients of similar expansion of $X(\mu,\lb)$.

Expanding $\BX_{11}$ in (\ref{4.2}) as above leads to
$$
        W_{2n,l}^{(k)}(t)\tau_{2n}(t,s)|_{s=-t}
        =\TT_{2n}(t)W_{2n,l}^{(k)}(t)\TT_{2n}(t).
$$
In particular, since $J_i^{(k)}$ ($k\le2$) and $W_{(n,i)}^{(k)}$
($k\le2$) are linear combinations of each other \cite{AvM3}:
\begin{align*}
W_{n,i}^{(0)}&=J_i^{(0)}=\dt_{i,0},\\
W_{n,i}^{(1)}&=J_i^{(1)}+nJ_i^{(0)},\\
W_{n,i}^{(2)}&=J_i^{(2)}+(2n-i-1)J_i^{(1)} + n(n-1)J_i^{(0)},
\end{align*}
one sees for $k=1$, 2 that
$$
J_i^{(k)}(t)\tau_{2n}(t,s)|_{s=-t}=\TT_{2n}(t)J_i^{(k)}\TT_{2n}(t).
$$
\end{proof}

Consider the following vertex operator\footnote{
        As before, $\BX$ treats $\TT$ as an extended vector
        $(\TT_n)_{n\in A}$, where $\TT_n\equiv0$ for $n$ odd.
        So $\chi(z)$ appearing in $\BX(z)$ acts on $\TT_n$ as multiplication
        by $z^n$, and $\Lb^\top$ acts on $\TT$ as $(\Lb^\top\TT)_n=\TT_{n-1}$.
        In practice, we always have even number of $\BX$'s acting on $\TT$,
        so there is always an even power of $\Lb^\top$, and $\TT_n$ for odd
        $n$ will never appear in our formulas.
}
$$
\BX(z):= \Lb^\top\BX_1(z) =
\Lb^\top e^{\sum t_iz^i}e^{-\sum\frac{z^{-i}}{i}\frac{\pl}{\pl t_i}}\chi(z),
$$
and define the kernel
$$
K_n(y,z):=\biggl(\frac{1}{\TT}\BX(y)\BX(z)\TT\biggr)_{2n}.
$$
It is easy to see that $(\BX(y)\BX(z)\TT)_{2n}=y\BX_1(y)\BX_1(z)\TT_{2n-2}$,
so by (\ref{4.3})
$$
K_n(y,z)=-\frac{(\BX_{12}(z,y)\tau)_{2n}}{\tau_{2n}}\biggr|_{s=-t}
=\frac{(\BX_{12}(y,z)\tau)_{2n}}{\tau_{2n}}\biggr|_{s=-t},
$$
and by (\ref{4.1})
$$
\biggl(\sum_{2m\le j<2n}\Psi_{1,j}(\mu)\Psi_{2,j}^*(\lb^{-1})\biggr)
\biggr|_{s=-t}
=K_n(\mu,\lb) - K_m(\mu,\lb).
$$
Here each term $\Psi_{1,j}(\mu)\Psi_{2,j}^*(\lb^{-1})$ on the left
hand side blows up along $s=-t$, but the poles from two successive
terms (for $j=2k$ and $j=2k+1$) cancel, as we saw earlier.

For $n\in A$, let
\begin{equation}
q_n(t,\lb):=
\begin{cases}
        \displaystyle
        \lb^{2m}\frac{\TT_{2m}(t-[\lb^{-1}])}{\sqrt{\TT_{2m}(t)\TT_{2m+2}(t)}}
        &\text{if $n=2m$},\\[12pt]
        \displaystyle
        \lb^{2m}\frac{(\pl/\pl t_1+\lb)\TT_{2m}(t-[\lb^{-1}])}
        {\sqrt{\TT_{2m}(t)\TT_{2m+2}(t)}}
        &\text{if $n=2m+1$}.
\end{cases}
\label{4.5}
\end{equation}
In the semi-infinite case, the $q_n$'s form a system of skew-orthogonal
polynomials \cite{AHvM}.

\begin{theorem}
The following holds:
\begin{equation}
\Pf(K_n(z_i,z_j))_{1\le i,j\le 2k}=
\Biggl(\frac{1}{\TT}\prod_{\substack{i=1\\\text{ordered}}}^{2k}\BX(z_i)\TT
\Biggr)_{2n},
\label{4.6}
\end{equation}
\begin{multline}
K_{n+1}(\mu,\lb)-K_n(\mu,\lb)
\\=e^{\sum_1^\iy t_i(\mu^i+\lb^i)}
\left(
q_{2n}(t,\lb)q_{2n+1}(t,\mu) - q_{2n}(t,\mu)q_{2n+1}(t,\lb)
\right),
\label{4.7}
\end{multline}
so in the semi-infinite case
\begin{equation}
K_N(\mu,\lb)= e^{\sum_1^\iy t_i(\mu^i+\lb^i)}
\sum_0^{N-1}\left(
q_{2n}(t,\lb)q_{2n+1}(t,\mu) - q_{2n}(t,\mu)q_{2n+1}(t,\lb)
\right).
\label{4.8}
\end{equation}
\end{theorem}

\begin{proof}
Using (\ref{PfaffFay}) and
\begin{align}
K_n(\mu,\lb)&=\biggl(\frac{\BX(\mu)\BX(\lb)\TT}{\TT}\biggr)_{2n}
\nonumber\\
&=(\mu-\lb)(\mu\lb)^{2n-2}e^{\sum_1^\iy t_i(\mu^i+\lb^i)}
\frac{\TT_{2n-2}(t-[\mu^{-1}]-[\lb^{-1}])}{\TT_{2n}(t)}\nonumber\\
&=(\lb^{-1}-\mu^{-1})(\mu\lb)^{2n-1}e^{\sum_1^\iy t_i(\mu^i+\lb^i)}
\frac{\TT_{2n-2}(t-[\mu^{-1}]-[\lb^{-1}])}{\TT_{2n}(t)},
\label{4.9}
\end{align}
the left hand side of (\ref{4.6}) becomes
$$
(z_1\cdots z_{2k})^{2n-1}e^{\sum_{j=1}^{2k}\sum_{i=1}^\iy t_iz_j^i}
\Dt(z^{-1})\frac{\TT_{2n-2k}(t-\sum_{j=1}^{2k}[z_j^{-1}])}{\TT_{2n}(t)}.
$$
This equals the right hand side of (\ref{4.6}), because
\begin{align*}
\lefteqn{
\left(\BX(z_{2k})\BX(z_{2k-1})\cdots\BX(z_2)\BX(z_1)\TT\right)_{2n}
}\quad&\\
&=z_{2k}^{2n-2}z_{2k-1}^{2n-2}\cdots z_1^{2n-2k}
e^{\sum t_i(z_1^i+\cdots+z_{2k}^i)}\\
&\qquad\cdot
\biggl[\prod_{1<j\le2k}\prod_{1\le i<j}\biggl(1-\frac{z_i}{z_j}\biggr)\biggr]
\TT_{2n-2k}\biggl(t-\sum_1^{2k}[z_i^{-1}]\biggr)\\
&=(z_1\cdots z_{2k})^{2n-1}\Dt(z^{-1})
e^{\sum t_i(z_1^i+\cdots+z_{2k}^i)}
\TT_{2n-2k}\biggl(t-\sum_1^{2k}[z_i^{-1}]\biggr).
\end{align*}
To prove (\ref{4.7}), we have
\begin{align*}
\lefteqn{
(\mu-\lb)\left(
(\mu\lb)^{2n}\frac
{\TT_{2n}(t-[\mu^{-1}]-[\lb^{-1}])}
{\TT_{2n+2}(t)}
-(\mu\lb)^{2n-2}\frac
{\TT_{2n-2}(t-[\mu^{-1}]-[\lb^{-1}])}
{\TT_{2n}(t)}
\right)
}&\\
&=(\mu-\lb)(\mu\lb)^{2n}\frac
{\TT_{2n}(t)\TT_{2n}(t-[\mu^{-1}]-[\lb^{-1}]) -
(\mu\lb)^{-2} \TT_{2n-2}(t-[\mu^{-1}]-[\lb^{-1}])\TT_{2n+2}(t)}
{\TT_{2n+2}(t)\TT_{2n}(t)}
\\
&=(\mu\lb)^{2n}\frac
{\{\TT_{2n}(t-[\mu^{-1}]),\TT_{2n}(t-[\lb^{-1}])\}
+(\mu-\lb)\TT_{2n}(t-[\mu^{-1}])\TT_{2n}(t-[\lb^{-1}])
}
{\TT_{2n+2}(t)\TT_{2n}(t)}\\
\intertext{using (\ref{3.3}),}
&=\biggl(
\lb^{2n}\frac{\TT_{2n}(t-[\lb^{-1}])}{\sqrt{\TT_{2n}(t)\TT_{2n+2}(t)}}
\frac{\mu^{2n}(\pl/\pl t_1+\mu)\TT_{2n}(t-[\mu^{-1}])}
{\sqrt{\TT_{2n}(t)\TT_{2n+2}(t)}}
- (\lb\leftrightarrow\mu)
\biggr)
\\
&=(q_{2n}(t,\lb)q_{2n+1}(t,\mu)-q_{2n}(t,\mu)q_{2n+1}(t,\lb))
\end{align*}
in terms of the skew-orthogonal polynomials (\ref{4.5}).  Multiplying this
with an exponential and noting (\ref{4.9}), one obtains (\ref{4.7}).
Summing up this telescoping sum yields (\ref{4.8}):
\begin{align*}
K_N(\mu,\lb)&=\biggl(\frac{\BX(\mu)\BX(\lb)\TT}{\TT}\biggr)_{2N}\\
&=(\mu-\lb)(\mu\lb)^{2N-2}e^{\sum_1^\iy t_i(\mu^i+\lb^i)}
\frac{\TT_{2N-2}(t-[\mu^{-1}]-[\lb^{-1}])}{\TT_{2N}(t)}\\
&=\sum_0^{N-1}e^{\sum_1^\iy t_i(\mu^i+\lb^i)}(q_{2n}(t,\lb)q_{2n+1}(t,\mu)
-q_{2n}(t,\mu)q_{2n+1}(t,\lb)).
\end{align*}
\end{proof}

\section{The exponential of the vertex operator maintains $\TT$-functions}

The purpose of this section is to show the following theorem:

\begin{theorem}For a Pfaffian $\TT$-function,
\be
\TT+a\BX(\lb)\BX(\mu)\TT
\ee
is again a Pfaffian $\TT$-function.
\end{theorem}

Remember that $\BX(\lb)\BX(\mu)$ acts on $\tilde \tau_{2n}(t)$, as
follows:
\be
\BX(\lb)\BX(\mu)\TT_{2n}(t)=
\left(1-\frac{\mu}{\lb}\right)\lb^{2n-2}\mu^{2n-1}e^
{\sum
t_{1}(\lb^i+\mu^i)}\TT_{2n-2}(t-[\lb^{-1}]-[\mu^{-1}]).
\ee
It is convenient to relabel $\TT_{2n}\rg\TT_{n}$
\medbreak

$\BX(\lb)\BX(\mu)\TT_{n}(t)=\displaystyle{\left(1-\frac{\mu}{\lb}\right)\lb^{2n-
2}\mu^{2n-1}
e^{\sum_{1}^{\iy}
t_{i}(\lb^i+\mu^i)}}\TT_{n-1}(t-[\lb^{-1}]-[\mu^{-1}])$
\begin{eqnarray}
&=&\frac{\mu}{\lb}(\lb-\mu)\left(\Lambda^{-1}e^{\sum_{1}^{\iy}
t_{i}(\lb^i+\mu^i)}e^{-\sum_{1}^{\iy}\left(\frac{\lb^{-i}+\mu^{-i}}{i}\right)
\frac{\pl}{\pl t_{i}}}
\chi(\lb^2\mu^2)\TT\right)_n \nonumber\\
&=:&\frac{\mu}{\lb}(\lb-\mu)\BY(\lb,\mu).
\end{eqnarray}
With this relabeling, the bilinear identity takes on the form
$$
\oint_{z=\iy}\tau_{n}(t-[z^{-1}])\tau_{m+1}(t'+[z^{-1}])e^{\sum_{1}^{\iy}
(t_{i}-t'_{i})z^i}z^{2n-2m-2}dz
$$
\be
+\oint_{z=0}\tau_{n+1}(t+[z])\tau_{m}(t'-[z])e^{\sum_{1}^{\iy}
(t'_{i}-t_{i})z^{-i}}z^{2n-2m}dz=0.
\ee

\begin{lemma} We have:
\begin{eqnarray*}
& &(1-\lb z)^{ -1}(1-\mu z)^{ -1}-\frac{1}{\lb\mu
z^2}\left(1-\frac{1}{\lb z}\right)^{ -1}\left(1-\frac{1}{\mu z}\right)^{
-1}\\
\hspace{2cm}&=&\frac{1}{\mu
-\lb}\delta\left(z-\frac{1}{\lb}\right)+\frac{1}{\lb-\mu}\delta\left(z-\frac{1}{
\mu}
\right).
\end{eqnarray*}
\end{lemma}

\proof See for instance \cite{AMSV}.

\begin{lemma}
\begin{eqnarray}
&
&\int_{z=\iy}\BY\tau_{n}(t-[z^{-1}])\tau_{m+1}(t'+[z^{-1}])e^{\sum_{1}^{\iy}
(t_i-t'_i)z^i}z^{2n-2m-2}dz \nonumber\\ &
&\qquad+\int_{z=0}\BY\tau_{n+1}(t+[z])\tau_{m}(t'-[z])e^{\sum_{1}^{\iy}
(t'_i-t_i)z^{-i}}z^{2n-2m}dz \nonumber\\
&=&\frac{1}{\mu-\lb}\Biggl(\mu^{2n}\lb^{2m}\tau_{n}(t-[\mu^{-1}])\tau_m(t'-[\lb^
{-1}]) e^{\sum_{1}^{\iy}(t'_i\lb^i+t_i\mu^{i})} \nonumber\\ &
&\hspace{2cm}-\lb^{2n}\mu^{2m}\tau_n(t-[\lb^{-1}])\tau_m(t'-[\mu^{-1}])
e^{\sum_{1}^{\iy}(t_i\lb^i+t'_i\mu^{i})}\Biggr) \nonumber\\
\end{eqnarray}
\end{lemma}

\begin{proof} Upon performing the following operations
$$
\left\{\begin{array}{l}
n\mapsto n-1\\
t\mapsto t-[\mu^{-1}]-[\lb^{-1}]\\
\mbox{multiplication by\,\,}
(\lb\mu)^{2n-1}e^{\sum^{\iy}_{1}t_{i}(\mu^i+\lb^i)}
\end{array}
\right.,
$$
the bilinear identity (5.3) yields
\begin{eqnarray*}
0&=&\oint_{z=\iy}~\tau_{n-1}(t-[z^{-1}]-[\mu^{-1}]-[\lb^{-1}])\tau_{m+1}
(t'+[z^{-1}])(\lb\mu)^{2n-2}\\ &
&\hspace{1cm}\left(1-\frac{z}{\lb}\right)
\left(1-\frac{z}{\mu}\right)
\frac{\lb\mu}{z^{2}}
~e^{\sum_{1}^{\iy}((t_i-t'_i)z^i+t_i(\mu^i+\lb^i))} z
^{2n-2m-2}dz\\
& & +\oint_{z=0}\tau_{n}(t+[z]-[\mu^{-1}]-[\lb^{-1}])\tau_{m}
(t'+[z])(\lb\mu)^{2n}
 \\
& &\hspace{1cm}\left(1-\frac{1}{\lb
z}\right)^{-1}\left(1-\frac{1}{\mu z}\right)^{-1}
\frac{1}{\lb\mu z^2}~e^{\sum_{1}^{\iy}((t'_i-t_i)z^{-i}+t_{i}(\mu^i+\lb^i))}z^{2n-2m}dz.
\end{eqnarray*}
Subtracting this expression (which is $=0$), the left hand side of
(5.5) equals

\begin{eqnarray*}
&
&\oint_{z=\iy}\tau_{n-1}(t-[z^{-1}]-[\lb^{-1}]-[\mu^{-1}])\tau_{m+1}
(t'+[z^{-1}]) \\ & &\hspace{3cm}e^{\sum_1^{\iy}\left((t_i-t'_i)z^i+
\left(t_i-\frac{z^{-i}}{i}\right)(\lb^i +\mu^i)\right)}
(\lb\mu)^{2(n-1)}z^{2n-2m-2}dz\\ &
&+\oint_{z=0}\tau_n(t+[z]-[\lb^{-1}]-[\mu^{-1}])\tau_m (t'-[z])\\ &
&\hspace{3cm}e^{\sum_1^{\iy}\left((t'_i-t_i)z^{-i}+
\left(t_i+\frac{z^i}{i}\right)(\lb^i +\mu^i)\right)}(\lb\mu)^{2n}z^{2n-2m}dz\\
&=&\oint_{z=\iy}\tau_{n-1}(t-[z^{-1}]-[\lb^{-1}]-[\mu^{-1}])\tau_{m+1}
(t'+[z^{-1}])e^{\sum_{1}^{\iy}
\left((t_i-t'_i)z^i+t_i(\lb^i+\mu^i)\right)}\\
& &\quad (\lb\mu )^{2n-2}\left(
\left(1-\frac{\lb}{z}\right)
\left(1-\frac{\mu}{z}\right)-
\frac{\lb\mu}{z^2}\left(1-\frac{z}{\lb}\right)
\left(1-\frac{z}{\mu}\right)\right)z^{2n-2m-2}dz\\
 & &+\oint_{z=0}\tau_{n}(t+[z]-[\lb^{-1}]-[\mu^{-1}])\tau_{m}
(t'-[z])e^{\sum_{1}^{\iy}\left((t'_i-t_i)z^{-i}+t_i(\lb^i
+\mu^i)\right)} (\lb\mu)^{2n}\\ & &\quad
\left((1-\lb z)^{-1}(1-\mu z)^{-1}-\frac{1}{\lb\mu
z^2}\left(1-\frac{1}{\lb z}\right)^{-1}
\left(1-\frac{1}{\mu z}\right)^{-1}\right)z^{2n-2m}dz\\
&=&\frac{1}{\mu-\lb}\Biggl(\mu^{2n}
\lb^{2m}\tau_{n}(t-[\mu^{-1}])\tau_{m}(t'-[\lb^{-1}])
e^{\sum_{1}^{\iy}(t'_i\lb^i+t_i\mu^i)}\\ & &\quad\quad ~~~~~~~~
-\lb^{2n}\mu^{2m}\tau_{n}(t-[\lb^{-1}])\tau_{m}(t'-[\mu^{-1}])
e^{\sum_{1}^{\iy}(t_i\lb^i+t'_i\mu^i)}\Biggr),
\end{eqnarray*}
ending the proof of the lemma.
\end{proof}

\medbreak

\noindent {\it Proof of theorem 5.1:} It suffices to prove
\begin{eqnarray*}
0&=&\oint_{z=\iy}(a+b\BY)\tau_{n}(t-[z^{-1}])(a+b\BY)\tau_{m+1}
(t'+[z^{-1}])e^{\sum_{1}^{\iy}(t_i-t'_i)z^i}z^{2n-2m-2}dz\\
& & +\oint_{z=0}(a+b\BY)\tau_{n+1}(t+[z])(a+b\BY)\tau_{m}
(t'-[z])e^{\sum_{1}^{\iy}(t'_i-t_i)z^{-i}}z^{2n-2m}dz.
\end{eqnarray*}
The coefficient of $a^2$ and $b^2$ vanishes, on view of the fact
that $\tau_{n}$ and $\BY\tau_{n}$ are Pfaffian $\tau$-functions. So it
suffices to show the vanishing of the $ab$-term.

\medbreak

\noindent $ab$-coefficient
\begin{eqnarray*}
&=&\oint_{z=\iy}\left(\BY\tau_{n}(t-[z^{-1}])\tau_{m+1}
(t'+[z^{-1}])+\tau_{n}(t-[z^{-1}])\BY\tau_{m+1}
(t'+[z^{-1}])\right)\\
& &\hspace{7cm}e^{\sum_{1}^{\iy}(t_i-t'_i)z^i}z^{2n-2m-2}dz\\
& &+\oint_{z=0}\left(\BY\tau_{n+1}(t+[z])\tau_{m}
(t'-[z])+\tau_{n+1}(t+[z])\BY\tau_{m}(t'-[z])\right)e^{\sum_{1}^{\iy}(t'_i-t_i)z
^{-i}}\\
& &\hspace{8cm}z^{2n-2m}dz.
\end{eqnarray*}
The first terms in each of the integrals can be evaluated by means
of lemma. The sum of the two terms equals
\begin{eqnarray}
 &&\frac{1}{\mu-\lb}\Biggl(\mu^{2n}\lb^{2m}
 \tau_{n}(t-[\mu^{-1}])\tau_{m}(t'-[\lb^{-1}])
  e^{\sum_{1}^{\iy}(t'_i\lb^i+t_{i}\mu^i)}\nonumber\\
   & &\quad
-\lb^{2n}\mu^{2m}\tau_{n}(t-[\lb^{-1}])\tau_m(t'-[\mu^{-1}])
e^{\sum_{1}^{\iy}(t_i\lb^i+t'_{i}\mu^i)}\Biggr).
\end{eqnarray}
Performing the exchange
$$
n\longleftrightarrow m,\quad t\longleftrightarrow t',\quad
z\longleftrightarrow z^{-1}
$$
gives an expression for the sum of the second terms in the
integrals; the sum of expression (5.6) and the same expression with
the exchange above is obviously zero.
\qed

\section{Examples}

\subsection{Symmetric matrix integrals}

Consider the matrix $m_n(t,s)$ of $(t,s)$-dependent moments,
\begin{equation}
\mu_{k\ell}(t,s):=\int\!\int_{\BR^2}x^k y^\ell
e^{\sum^\iy_1(t_ix^i-s_iy^i)} F(x,y)dx\,dy,\quad t,s \in\BC^\iy,
\label{5.1}
\end{equation}
with regard to a skew-symmetric weight $F(x,y)$, satisfying $$
F(x,y)=-F(y,x). $$
 Then
\begin{align}
\tau_n(t,s)
&:=\det m_n (t,s)
\label{5.2}
\\
&=\int\!\cdots\!\int_{\BR^{2n}}\prod^n_{k=1}
\left(e^{\sum^\iy_{i=1}(t_i
x_k^i-s_iy^i_k)} F(x_k,y_k)\right)
\Dt_n(x)\Dt_n(y)\vec{dx}\,\vec{dy}.
\label{5.3}
\end{align}

The proof that the determinant (\ref{5.2}) of the moment matrix
equals the multiple integral (\ref{5.3}) is based on an identity
involving Vandermonde determinants and can be found in \cite{AvM2}. Then
$$
\TT_{2n}(t):=\Pf m_{2n}(t,-t)=\sqrt{\tau_{2n}(t,-t)}
$$
satisfies equations (0.6) up to (0.8).

Moreover, the moments $\mu_{ij}$ in (\ref{5.1}) satisfy the equations
$$
\frac{\pl\mu_{ij}}{\pl t_k}=\mu_{i+k,j}\quad\mbox{and}\quad
\frac{\pl\mu_{ij}}{\pl s_k}=-\mu_{i,j+k},
$$
and so $m:=m_\iy$ satisfies (\ref{1.1}).

The skew-symmetry of $F$ above implies the skewness of
$m_\iy(0,0)$; so, by theorem 2.1, we have
$$
\mu_{ij}(t,s)=-\mu_{ji}(-s,-t).
$$
Therefore also, $m_\iy$, $S_1$, $S_2$, $\Psi_1$, $\Psi_2$, $h$ and $\tau$
have the properties mentioned in theorem 2.1 and the vector
$(\tau_n)_{n\geq 0}$ is a 2-Toda $\tau$-vector, i.e. the
corresponding wave vectors $\Psi_i$ and the matrices $L_i$ satisfy
the formulae of theorem 2.1.

For skew-symmetric weights $F(x,y)$ of the special form
$$
F(x,y):=e^{V(x)+V(y)}I_E(x)I_E(y)\operatorname{sign}(x-y),\quad\mbox{for
an interval $E\subset\BR$},
$$
 and for a
union of intervals $E \subset \BR$, the expression
$\TT_{2n}=\tau_{2n}(t,-t)^{1/2}$ equals the integral over symmetric
matrices, given in
$$
\TT_{2n}(t)=\sqrt{\tau_{2n}(t,-t)}
=\int_{{\cal S}_{2n}(E)}e^{\Tr(V(X)+\sum_1^\iy t_iX^i)}
dX,
$$
 for the Haar measure $dX$ on symmetric matrices and
$$
{\cal S}_{2n}(E):=\{\mbox{$2n\times 2n$ symmetric matrices $X$
with spectrum $\subset E$}\}.
$$

 In \cite {AvM5}, we worked out the Virasoro
constraints satisfied by (\ref{0.15}), which then leads to
inductive expressions for those integrals, involving
Painlev\'e-like expressions.

\subsection{Quasiperiodic solutions}

In this subsection, we shall combine the construction of quasi-periodic
solutions of 2-Toda lattice \cite{mfd,UT} and the theory of Prym varieties
\cite{mfd:p} to obtain quasiperiodic solutions of the Pfaff lattice.
While we put stress on the semi-infinite case in the present paper,
this gives a non-trivial example in the bi-infinite case.

A 2-Toda quasiperiodic solution is given by some deformation of a line
bundle $\LR$ on a complex curve (Riemann surface) $C$, with the time
variables playing the role of deformation parameters, so the orbit
under the 2-Toda flows is parametrized by the Jacobian of $C$.
If $C$ is equipped with an involution $\iota\colon C\to C$, and if
$\LR$ satisfies a suitable antisymmetry condition with respect to
$\iota$, then the 2-Toda flows can be restricted to preserve the
antisymmetry of $\LR$, giving a solution of Pfaff lattice.
The Prym variety $P$ of $(C,\iota)$ naturally appears as the
restricted parameter space.  The vanishing of every other
$\tau_n(t,-t)$ (see (\ref{0.4}) or (\ref{2.6})) indicates that
the space of $\LR$'s which satisfy the antisymmetry condition must
consist of two connected components, $P$ and $P^-$.
This means the involution $\iota$ has no fixed points.
So, in general a quasiperiodic solution of the Pfaff lattice does not
satisfy the BKP equation and vice versa, since the orbit of a
quasiperiodic solution of the BKP equation is isomorphic to the Prym
variety of a curve with involution having at least two fixed points.

\subsubsection*{Preliminary on the geometry of curves}
A line bundle on a complex curve $C$ is defined by a divisor
$D=\sum m_ip_i$, $m_i\in\BZ$, $p_i\in C$, i.e., a set of points
$p_i$ on $C$ with (positive or negative) multiplicities $m_i$, as
$\LR=\OR(D)$.  Its local sections (on an open set $U\subset C$, say)
are meromorphic functions on $U$ which have poles of order at most
$m_i$ (zeros of order at least $-m_i$) at $p_i$.
The number $d:=\sum m_i$ is called the degree of $\LR$.
For $\LR=\OR(D)$ and $m$,~$n\in\BZ$, $p$,~$q\in C$, we denote
$\LR(mp+nq)=\OR(D+mp+nq)$ etc.
A deformation of $\LR$ can be described as a deformation of $D$,
like $D_{t,s}=\sum m_ip_i(t,s)$, but in the 2-Toda theory it is
more convenient to describe it by requiring its local sections
to have some exponential behaviors at prescribed points, as we
shall see later.

Two line bundles $\OR(D_1)$ and $\OR(D_2)$ are isomorphic if the
divisors $D_1$ and $D_2$ are ``linearly equivalent,'' i.e., if
they differ by the divisor of a global meromorphic function on $C$.
Jacobian $J$ of $C$ is the space (Lie group) of isomorphism classes of
degree 0 line bundles on $C$.  It becomes a principally polarized
abelian variety of dimension $g:=\text{genus of }C$, i.e., $J$ is
a complex torus $\BC^g/\Gamma$, $\BC^g\supset\Gamma\simeq\BZ^{2g}$,
for which there is a divisor (codimension 1 subvariety)
$\Theta\subset J$, such that some positive integer multiple of
$\Theta$ defines an embedding of $J$ into a complex projective
space, and $\Theta$ is ``rigid'' in the sense that it has
no deformation in $J$ except parallel translations.
A complex torus $\BC^g/\Gamma$ is a principally polarized abelian variety
if and only if, after some change of coordinates by $GL(g,\BC)$,
the lattice $\Gamma$ becomes $\BZ^g+\Omega\BZ^g$ for some complex symmetric
$g\times g$ matrix $\Omega$ with positive definite imaginary part.
On a principally polarized abelian variety $\BC^g/(\BZ^g+\Omega\BZ^g)$,
there is a special quasiperiodic function (i.e., holomorphic function
on $\BC^g$ that satisfies some quasiperiodicity condition with respect
to $\BZ^g+\Omega\BZ^g$) called Riemann's theta function $\vartheta$,
defined by
$$
\vartheta(z)=\sum_{m\in\BZ^g}\exp(2\pi im^tz+\pi im^t\Omega m).
$$
The theta divisor $\Theta$ becomes the zero divisor of $\vartheta$.

If $C$ has a (holomorphic) involution $\iota\colon C\to C$ (i.e.,
$\iota^2=\text{id}$), $J$ gets an involution $\iota^*$ induced by
$\iota$.  The Jacobian $J'$ of the quotient curve $C'=C/\iota$, and
the Prym variety $P$ of the pair $(C,\iota)$ (or $(C,C')$) appear
in $J$ roughly as the $\pm1$ eigenspaces of $\iota$: $\tilde
J'=J'/($some~subgroup of order~2$)\subset J$ and $P\subset J$ are
subabelian varieties of $J$, such that $\iota|_{\tilde J'}=+1$,
$\iota|_P=-1$, and
$J\simeq(J'\times P)/(\text{finite subgroup})$.  When $\iota$ has
at most two fixed points, the restriction of $\Theta$ on $P$ gives
twice some principal polarization on $P$ (the restriction $\vartheta|_P$
becomes the square of the Riemann theta function on $P$ defined by
this polarization).

\subsubsection*{Quasiperiodic solutions of 2-Toda lattice}
Let $C$ be a nonsingular complete curve on $\BC$ (compact Riemann surface)
of genus $g$, let $\LR$ be a line bundle of degree $g-1$ on $C$, let
$p$,~$q\in C$ be distinct points.  Let us choose local coordinates
$z^{-1}$ at $p$ and $z$ at $q$, and trivializations of $\LR(p)$ at
$p$ and $q$,
$$
\sigma_p\colon\LR_p(p)\simeq\OR_p\quad\text{and}\quad
\sigma_q\colon\LR_q\simeq\OR_q.
$$
For $t$,~$s\in\BC^\infty$, let $\LR_{t,s}$ be the line bundle whose
(local holomorphic) sections are (local holomorphic) sections of $\LR$
away from $p$ and $q$, and at $p$ (resp.\ $q$) have singularities of
the form $e^{\sum t_iz^i}\cdot(\text{holomorphic})$
(resp.\ $e^{\sum s_iz^{-i}}\cdot(\text{holomorphic})$).
For ``generic''\footnote{
        Here generic means that $\Gamma(\LR_{t,s}(np-nq))=\{0\}$ holds.
        For a degree $g-1$ line bundle $\LR$, this condition holds for
        almost all $(n,t,s)\in\BZ\times\BC^\infty\times\BC^\infty$,
        and implies that $\dim\Gamma(\LR_{t,s}((n+1)p-nq))=1$.
}
$(n,t,s)\in\BZ\times\BC^\infty\times\BC^\infty$, the wave functions
$\Psi_{1,n}$, $\Psi_{2,n}$ are obtained from a (unique) section
$\varphi_n(t,s)$ of
$$
\LR_{t,s}((n+1)p-nq),
$$
which has the form $z^ne^{\sum t_iz^i}(1+O(z^{-1}))$ at $p$ via $\sigma_p$,
i.e.,
\begin{equation}
\begin{aligned}
\Psi_{1,n}(t,s,z)&:=\sigma_p(\varphi_n(t,s))=z^ne^{\sum t_iz^i}(1+O(z^{-1})),
\\
\Psi_{2,n}(t,s,z)&:=\sigma_q(\varphi_n(t,s))
=z^ne^{\sum s_iz^{-i}}(h_n(t,s)+O(z)).
\end{aligned}
\end{equation}
The adjoint wave functions
\begin{align*}
\Psi^*_{1,n}&=z^{-n}e^{-\sum t_iz^i}(1+O(z^{-1})),
\\
\Psi^*_{2,n}&=z^{-n}e^{-\sum s_iz^{-i}}(h_n(t,s)^{-1}+O(z))
\end{align*}
are defined similarly, by using
$$
(\LR_{t,s})^*(-np+(n+1)q)=(\LR^*)_{-t,-s}(-np+(n+1)q),
$$
in place of $\LR_{t,s}((n+1)p-nq)$, where we denote
$$
\LR^*:={\cal H}om(\LR,\omega)=\LR^{-1}\otimes\omega,
$$
with $\omega$ being the dualizing sheaf (the canonical bundle,
i.e., the line bundle of holomorphic 1-forms), and,
in place of $\sigma_p$ and $\sigma_q$, trivializations
$$
\sigma_p^*\colon\LR_p^*\simeq\OR_p\quad\text{and}\quad
\sigma_q^*\colon\LR_q^*(q)\simeq\OR_q,
$$
for which the maps
\be
\begin{aligned}
\LR_p(p)\otimes\LR^*_p\ni(\phi,\psi)&\mapsto\sigma_p(\phi)\sigma_p^*(\psi)dz/z
\in\omega(p)_p,\\
\LR_q\otimes\LR^*_q(q)\ni(\phi,\psi)&\mapsto\sigma_q(\phi)\sigma_q^*(\psi)dz/z
\in\omega(q)_q
\end{aligned}
\label{5.4}
\ee
extend to the canonical map
$$
\LR(p)\otimes\LR^*(q)\stackrel{\simeq}{\to}\omega(p+q).
$$
Hence for general $(n,t,s)$, $(m,t',s')\in\BZ\times\BC^\infty\times\BC^\infty$,
$$
\Psi_{i,n}(t,s,z)\Psi_{i,m}^*(t',s',z)dz/z,\quad i=1,\ 2
$$
become expansions at $p$ and $q$, respectively, of a holomorphic
1-form on $C\setminus\{p,q\}$, so by the residue calculus the pair
$\Psi$, $\Psi^*$ satisfies the bilinear identities (\ref{1.7}).

\subsubsection*{Quasiperiodic solutions of Pfaff lattice}
In the above construction, suppose $C$ has an involution
$\iota\colon C\to C$ with no fixed point.  In this case $g$ is
odd, $g=2g'-1$, with $g'$ being the genus of the quotient curve
$C'=C/\iota$. Suppose $q=\iota(p)$, and $\LR$ satisfies
\begin{equation}
\iota^*(\LR)\simeq\LR^*,\quad\text{so that}\quad
\LR\otimes\iota^*\LR\simeq\omega.
\label{antisym}
\end{equation}
Choose the local coordinates $z^{\mp1}$ and the trivializations
$\sigma_p$, $\sigma_q$, $\sigma_p^*$, $\sigma_q^*$
at $p$ and $q=\iota(p)$, such that $z\cdot\iota^*z\equiv1$ and
$\sigma_q=\iota^*\circ\sigma^*_p\circ\iota^*$ hold. (We then have
$\sigma_q^*=-\iota^*\circ\sigma_p\circ\iota^*$, with the minus
sign due to the fact that $dz/z$, which appear in (\ref{5.4}),
satisfy $\iota^*(dz/z)=-dz/z$.)
Then the wave and adjoint wave functions constructed above satisfy
(\ref{2.3}), so they lead to a quasiperiodic solution of the Pfaff
lattice when $s=-t$ (and skipping every other $n$).

The orbit of the 2-Toda flows is parametrized by the Jacobian $J$ of $C$,
and the $\tau$-functions are written in terms of Riemann's theta function
of $J$.  The orbit of the Pfaff flows will become the Prym variety $P$
of $(C,\iota)$, with $\tilde\tau$ given by the Prym theta function.
To be more precise, let $J_{g-1}$ be the moduli space of the
isomorphism classes of line bundles of degree $g-1$ on $C$.  This is a
principal homogeneous space\footnote{
        Hence $J_{g-1}$ is (non-canonically) isomorphic to $J$.
        We choose this isomorphism in such a way that
        $\Theta\subset J_{g-1}$ is identified with the zero locus
        of Riemann's theta function for $J$.
}
over $J$, on which the theta divisor
$$
\Theta:=\bigl\{\LR\in J_{g-1}\bigm|\Gamma(\LR)\ne(0)\bigr\}
$$
is canonically defined. The set of $\LR\in J_{g-1}$ satisfying
(\ref{antisym}) becomes the disjoint union $P_{g-1}\cup P_{g-1}^-$,
where
\begin{align*}
P_{g-1}&:=\bigl\{\LR\in J_{g-1}\bigm|\text{$\LR$ satisfies
(\ref{antisym}) and $\dim \Gamma(\LR)$ is even}\bigr\},
\\
P_{g-1}^-&:=\bigl\{\LR\in J_{g-1}\bigm|\text{$\LR$ satisfies
(\ref{antisym}) and $\dim \Gamma(\LR)$ is odd}\bigr\}
\end{align*}
are principal homogeneous spaces over the Prym $P$. We have
$$
P_{g-1}^-\subset\Theta\quad\text{and}\quad P_{g-1}\cdot\Theta=2\Xi,
$$
for some divisor $\Xi\subset P_{g-1}$ which gives a principal
polarization on $P_{g-1}$.  Since $\Theta$ is the zero locus of
Riemann's theta function $\vartheta$ of the Jacobian $J$, this
means $\vartheta$ vanishes identically on $P_{g-1}^-$, and the
restriction $\vartheta|_{P_{g-1}}$ becomes the square of Riemann's
theta function $\vartheta_P$ of $(P,\Xi)$, which is called the
Prym theta function.

For a 2-Toda quasiperiodic solution, the discrete time flow
(shift of $n$ by 1) is given by the shift $\LR\mapsto\LR(p-q)$.
In the present case, since $q=\iota(p)$, this flow preserves
condition (\ref{antisym}).  Moreover, we have
\begin{equation}
\forall p\in C,\;\forall\LR\in J_{g-1}\colon\left\{
\begin{array}{rcl}
\LR\in P_{g-1} &\Rightarrow & \LR(p-\iota(p))\in P_{g-1}^-, \\
\LR\in P_{g-1}^- &\Rightarrow & \LR(p-\iota(p))\in P_{g-1}.
\end{array}
\right.
\label{alternate}
\end{equation}
so that $\LR(np-n\iota(p))$'s alternate between $P_{g-1}$ and $P_{g-1}^-$,
and every other $\tau$ function vanishes identically when $s=-t$.
Shifting the discrete index $n$ by 1 if necessary, we may assume that
$\tau_n(t,s)$ satisfies (\ref{0.4}) or (\ref{2.6}).

\subsubsection*{Explicit formulas}
Explicit formulas for $\Psi$, $\Psi^*$ and $\tau$ can be given in
terms of Riemann's theta function for $J$, and hence explicit formulas
for $\TT$ can be given in terms of the Prym theta function for $P$.

Taking a basis $A_i$, $B_i$ ($i=1,\dots,g$) of $H_1(C,\BZ)$ such that
$A_i\cdot B_j=\dt_{i,j}$ and $A_i\cdot A_j=B_i\cdot B_j=0$,
let $\omega_i$ ($i=1,\dots,g$) be a basis of the space of holomorphic
1-forms such that
$$
\int_{A_i}\omega_j=\delta_{i,j}.
$$
Then
$$
\int_{B_i}\omega_j=\Omega_{i,j}
$$
gives a complex symmetric matrix $\Omega$ with positive definite
imaginary part, and $J=\BC^g/(\BZ^g+\Omega\BZ^g)$ becomes the Jacobian
of $C$.  Choosing a point $p\in C$, the map
$$
\alpha\colon
C\ni x\mapsto\biggl(\int_p^x\omega_1,\dots,\int_p^x\omega_g\biggr)
\in J
$$
is well-defined and gives an embedding of $C$ into $J$.
Composing $\alpha$ with a translate of Riemann's theta function:
\begin{equation}
q(x):=\vartheta(\alpha(x)+a),\quad a\in\BC^g,
\label{theta-part}
\end{equation}
one obtains a multi-valued function on $C$ which is single-valued
around the $A$-cycles.

Next, let $\zeta^{(p)}_n$, $n=1$,~2,\dots, be the differentials of
the second kind (meromorphic 1-forms with no residues) with poles
only at $p$ of the form $d(z^n+O(1))$ and no $A$-periods
($\int_{A_i}\zeta^{(p)}_n=0$), and let $\zeta^{(q)}_n$, $n=1$,~2,\dots,
be defined similarly, with $p$ replaced by $q$ and $z$ by $z^{-1}$
(recall that $z^{-1}$ (resp.\ $z$) is the local coordinate at $p$
(resp.\ $q$)). Let $\zeta_0$ be the differential of the third kind
(meromorphic 1-form with simple poles) with no $A$-periods and
poles only at $p$ and $q$ of the form $dz/z+O(1)$.  Then, given
$(n,t,s)\in\BZ\times\BC^\infty\times\BC^\infty$, the multi-valued
holomorphic function
\begin{equation}
C\ni x\mapsto\varepsilon(x):=\exp\biggl(
\int^x\biggl(n\zeta_0+\sum_{n=1}^\infty t_n\zeta^{(p)}_n
+\sum_{n=1}^\infty s_n\zeta^{(q)}_n\biggr)
\biggr)
\label{exp-part}
\end{equation}
has singularities at $p$ and $q$ of the form $z^ne^{\sum t_nz^n}$
and $z^ne^{\sum s_nz^{-n}}$, respectively, and is single-valued
around $A$-cycles.  The product of the form $\varepsilon(x)q(x)/q(p)$,
where $\varepsilon(x)$ and $q(x)$ are as in (\ref{theta-part}) and
(\ref{exp-part}), with
\begin{equation}
a=a(n,t,s)=n\alpha(q)+\sum_{i=1}^\infty t_iU_i+\sum_{i=1}^\infty s_iV_i+a_0,
\quad\forall a_0\in\BC^g,
\label{5.10}
\end{equation}
and $U_i=-(d/d(z^{-1}))^i\alpha(p)/(i-1)!$, $V_i=-(d/dz)^i\alpha(q)/(i-1)!$,
gives function on $(n,t,s)$, and hence the wave functions $\Psi$, with
the desired properties:
\begin{itemize}
\item $\varphi_n(t,s;x)$ is single-valued around the $A$-cycles, and
when $x$ goes around $B_i$, it is multiplied by a factor independent of
$(n,t,s)$,
\item $\varphi_n(t,s;x)\simeq z^ne^{\sum t_iz^i}(1+O(z^{-1}))$
at $x\simeq p$, and\\
$\varphi_n(t,s;x)\simeq z^ne^{\sum s_iz^{-i}}(h_n(t,s)+O(z))$
at $x\simeq q$.
\end{itemize}
The adjoint wave functions $\Psi^*$ are obtained similarly, from
$\varepsilon(x)^{-1}\vartheta(\alpha(x)-a)/\vartheta(-a)$ with the same
$a$ as above.

The 2-Toda $\tau$-function can be computed from those formulas as
\begin{equation}
\tau_n(t,s)=\exp(Q(n,t,s))\vartheta(a(n,t,s))
\label{5.11}
\end{equation}
for some quadratic form $Q(n,t,s)$, i.e.,
$$
Q(n,t,s)=\sum_{i,j=1}^\infty Q_{i,j}t_it_j
+\sum_{i,j=1}^\infty Q'_{i,j}s_is_j
+\sum_{i=1}^\infty n(q_it_i+q'_is_i),
$$
with $Q_{i,j}=Q_{j,i}$ appearing in the Laurent expansion
of the integral of $\zeta^{(p)}_i$ or $\zeta^{(p)}_j$ as
$$
\int^x\zeta^{(p)}_i=z^i-2\sum_{j=1}^\infty Q_{i,j}z^{-j}/j
\quad\text{for}\quad x\simeq p,
$$
$Q'_{i,j}=Q'_{j,i}$ appearing similarly in the Laurent expansion of
the integral of $\zeta^{(q)}_i$ or $\zeta^{(q)}_j$ as
$$
\int^x\zeta^{(q)}_i=z^{-i}-2\sum_{j=1}^\infty Q'_{i,j}z^j/j
\quad\text{for}\quad x\simeq q,
$$
and $q_i$ and $q'_i$ appearing similarly in the expansions
$$
\int^x\zeta_0=\log z-\sum_{j=1}^\infty q_jz^{-j}/j
\quad\text{for}\quad x\simeq p
$$
and
$$
\int^x\zeta_0=\log z-\sum_{j=1}^\infty q'_jz^j/j
\quad\text{for}\quad x\simeq q.
$$

Suppose $C$ has an involution $\iota$ with no fixed points, so that
$g=2g'-1$ with $g'$ being the genus of the quotient curve $C'=C/\iota$.
Suppose $q=\iota(p)$.
Take the cycles $A_i$, $B_i$ ($i=1,\dots,g$) in such a way that
$\iota(A_i)\simeq A_{g+1-i}$, $\iota(B_i)\simeq B_{g+1-i}$.
Then $\iota^*(\omega_i)=\omega_{g+1-i}$, and $\Omega$ satisfies
$\Omega_{i,j}=\Omega_{g+1-i,g+1-j}$.  The map
$\tilde\iota\colon\BC^g\ni(z_1,\dots,z_g)\mapsto(z_g,\dots,z_1)\in\BC^g$
maps the lattice $\Gamma:=\BZ^g+\Omega\BZ^g$ onto itself, and the embeddings
$$
\tilde J'=J'/(\BZ/2\BZ)\subset J\quad\text{and}\quad P\subset J
$$
are given by the images under $\pi_L\colon\BC^g\to\BC^g/\Gamma$
of the $\pm1$-eigenspaces of $\tilde\iota$: Setting
$$
R':=(\delta_{i,j}+\delta_{i,g+1-j})_{1\le i\le g,1\le j\le g'}
\quad\text{and}\quad
R'':=(\delta_{i,j}-\delta_{i,g+1-j})_{1\le i\le g,1\le j\le g'-1},
$$
so that $\BC^g_+:=R'\BC^{g'}$ and $\BC^g_-:=R''\BC^{g'-1}$ are the
$\pm1$-eigenspaces of $\tilde\iota$, and for any $z\in\BC^g$,
$z':=(1/2)(R')^t z$ and $z'':=(1/2)(R'')^t z$ give the decomposition
$z=R'z'+R''z''\in\BC^g_+\oplus\BC^g_-$, we have
\begin{align*}
\tilde J'=\BC^{g'}/(\varepsilon\BZ^{g'}+\Omega'\BZ^{g'})
&\simeq \pi_L(R'\BC^{g'})\subset \BC^g/(\BZ^g+\Omega'\BZ^g)\\
z'&\mapsto R'z'
\end{align*}
and
\begin{equation}
\begin{aligned}
P=\BC^{g'-1}/(\BZ^{g'-1}+\Omega''\BZ^{g'-1})
&\simeq  \pi_L(R''\BC^{g'-1})\subset\BC^g/(\BZ^g+\Omega'\BZ^g)\\
z''& \mapsto R''z'',
\end{aligned}
\label{5.12}
\end{equation}
where $\varepsilon=\diag(1,1,\dots,1,1/2)$,
$$
\Omega'=\biggl(
\frac{\Omega_{i,j}+\Omega_{i,g+1-j}}{(1+\delta_{i,g'})(1+\delta_{j,g'})}
\biggr)_{1\le i,j\le g'}
\quad\text{and}\quad
\Omega''=(\Omega_{i,j}-\Omega_{i,g+1-j})_{1\le i,j\le g'-1}.
$$
In (\ref{5.10}), suppose $a_0=R''a''_0\in R''\BC^{g'-1}$.
Since, by definition, $\alpha(q)=\alpha(q)-\alpha(p)\in\pi_L(\BC^g_-)$
and $\tilde\iota(U_i)=V_i$, we then have $a(n,t,-t)=R''a''(n,t)$,
where
$$
a''(n,t)=\frac{1}{2}(R'')^t a(n,t,-t)
=(R'')^t\Bigl(\frac{1}{2} n\alpha(q)+\sum_{i=1}^\infty t_i U_i\Bigr)+a''_0.
$$
Hence by using (\ref{5.11}) and (\ref{5.12}), and noting that
$Q'_{i,j}=Q_{i,j}$ and $q_i'=-q_i$, we have
$$
\TT(t)=\exp(\tilde Q(n,t))\vartheta_P(a''(n,t)),
$$
where
$$
\tilde Q(n,t)=\sum_{i,j=1}^\infty Q_{i,j}t_it_j+\sum_{i=1}^\infty q_int_i,
$$
and
$$
\vartheta_P(z)=\sum_{m\in\BZ^{g'-1}}\exp(2\pi i m^tz+\pi im^t\Omega''m),
\quad\text{for}\quad z\in\BC^{g'-1}.
$$

\end{document}